\documentclass[11pt,a4paper]{article}
\usepackage{jheppub}
\usepackage[FIGTOPCAP]{subfigure}
\usepackage{tensor}

\renewcommand{\Im}{\mathrm{Im}}
\renewcommand{\Re}{\mathrm{Re}}

\title{Holographic plasma and anyonic fluids}
\author[a,b]{Daniel K.~Brattan,\note{E-mail address: danny.brattan@gmail.com}}
\author[b]{Gilad Lifschytz\note{E-mail address: giladl@research.haifa.ac.il}}
\affiliation[a]{Physics Department, Technion- Israel Institute of Technology,
Technion City - Haifa, \\ 32000, Israel.}
\affiliation[b]{Department of Mathematics-Physics-Computer Science, University
of Haifa at Oranim, \\ Qiryat Tivon, 36006, Israel.}

\abstract{We use alternative quantisation of the $D3/D5$ system to explore properties of a strongly coupled charged plasma and strongly coupled anyonic fluids.
The $S$-transform of the $D3/D5$ system is used as a model for charged matter interacting with a $U(1)$ gauge field in the large coupling regime, and we compute the dispersion relationship of the propagating electromagnetic modes as the density and temperature are changed. A more general $SL(2,\mathbb{Z})$ transformation gives a strongly interacting anyonic fluid, and we study its transport properties as we change the statistics of the anyons and the background magnetic field.}

\begin{document}

\maketitle
\flushbottom

\section{Introduction}
\label{background}

{\ Much work has been done in recent years exploring strongly coupled CFTs with $U(1)$ global symmetry and their associated conserved currents via the gauge/gravity relationship. These systems have a rich variety of physically relevant phenomena to condensed matter, such as superfluidity, non-Fermi liquids, fractional quantum Hall effect etc. Less explored are strongly coupled CFTs with a local $U(1)$ gauge symmetry. Such theories can also be tackled using the gauge/gravity duality if they live in $(2+1)$ dimensions. In fact, as shown in \cite{Witten:2003ya}, given any CFT with a global $U(1)$ charge there is a well defined procedure generated by an element of $SL(2,\mathbb{Z})$ to turn it into a CFT with a local $U(1)$ gauge symmetry, which also has a conserved current. An example of a phenomenologically relevant model with such a conformal gauge field is given by the large charge limit of zero temperature $(2+1)$-dimensional QED\footnote{In condensed matter physics one application of QED$_{2+1}$ and its 
cousins can be seen in the two-dimensional $SU(N_{f})$ Heisenberg spin model (and variants) on a square lattice where the effective Hamiltonian turns out to be the fermionic term of lattice QED$_{2+1}$\cite{PhysRevB.37.3774}.}.}

{\ Given a holographic description of a CFT with a global $U(1)$ current one can find a holographic description of the $SL(2,\mathbb{Z})$ transformed field theory and compute correlation functions of the transformed conserved current. Thus we can describe using holographic techniques field theories whose field content has some matter coupled to a dynamical gauge field (in the sense that the gauge field is integrated over in the path integral) but with no Maxwell term. We can however add a Chern-Simons' term by an appropriate $SL(2,\mathbb{Z})$ transformation. Using such gravitational duals we can begin to study properties of these theories at finite temperature and density, i.e.~we can study strongly coupled ``real'' plasma physics in $(2+1)$ dimensions.}

{\ A second application of $(2+1)$-dimensional gauge theories in condensed matter physics concerns the appearance of ``anyons'', particles whose statistics interpolate between Bose-Einstein and Fermi-Dirac, which may be useful in understanding the quantum Hall effect. A simple model of such systems is given by a Chern-Simon's gauge theory with charged matter. The sourced Chern-Simon's equations of motion, with level $K$, attach magnetic flux to an electrically charged particle. Hence a point particle with charge $e$ sourcing the current also carries $(2\pi)/K$ units of magnetic flux. Moving one such point particle around another leads to an Aharanov-Bohm phase $ \Delta \theta = \frac{e^2}{K}$ where $\Delta \theta$ is the exchange phase angle. We see that it is Chern-Simon's level dependent and potentially neither integer nor half-integer. This leads to the unusual statistics of these anyonic particles \cite{PhysRevLett.49.957,Arovas:1985yb}.  A holographic model for an anyonic superconductor which bears 
some resemblence to our system has recently been investigated \cite{Jokela:2013hta}.}

\subsection{$SL(2,\mathbb{Z})$ transformation of a CFT}
\label{alternativequantisation}

{\ Let us start with a $(2+1)$-dimensional CFT which has a conserved current $J^{\mu}$ that can couple to a background source $A_{\mu}$. We can define another CFT also with a conserved current by an $SL(2,\mathbb{Z})$ transformation of the original theory using the prescription we shall now review \cite{Witten:2003ya,Leigh:2003ez,Herzog:2007ij}. First note that $SL(2,\mathbb{Z})$ is generated by an $S$-transformation and a $T$-transformation. The $S$-transformed CFT is obtained by including the background field ${A}_{\mu}$ in the path integral, but with no kinetic term.  A new conserved current in this CFT is $\frac{1}{2\pi}\int \epsilon _{\mu \nu\rho}\partial^{\nu}A^{\rho}$, which can be coupled to some new background vector field $C_{\mu}$, thus defining a new generating functional. A $T$-transformation is induced by adding to the theory a Chern-Simon term for the background field. The $S$-transformation and the $T$-transformation do not commute. The $U(1)$ current in the CFT and the Hodge dual of the  
background field strength transform as a 
doublet under the $SL(2,\mathbb{Z})$ transformation.}

{\ Given the above we can readily determine the transformation of the two point function between the CFTs. Assuming $SO(2)$ rotational invariance, the general expression for the current-current correlator in a finite temperature CFT is 
  \begin{eqnarray}
    \label{Eq:GenericCurrentCorrelator}
    \left\langle {J}_{\mu}(p) {J}_{\nu}(-p) \right\rangle
  &=& \sqrt{p^2} \left[ {C}_{(\mathrm{L})}(p) P^{(L)}_{\mu \nu} + {C}_{(\mathrm{T})}(p) P^{(T)}_{\mu \nu} \right] + {W}(p) \Sigma_{\mu \nu} 
  \end{eqnarray}
where the functions ${C}_{(\mathrm{T})}$, ${C}_{(\mathrm{L})}$ and ${W}$ are scale invariants and the tensor structures displayed above are defined in terms of the Minkowski metric $\eta_{\mu \nu}$, spatial metric $\delta_{ij}$, momentum $p_{\mu} = \left(\omega, k_{i}\right)$ and $p^2 = - \omega^2 + \vec{k}^2$ by 
  \begin{eqnarray}
   P_{\mu \nu} = \eta_{\mu \nu} - p_{\mu} p_{\nu} /p^2 - P_{\mu \nu}^{(T)} \; ,  P^{(T)}_{tt} = P^{(T)}_{ti} = 0 \; , P^{(T)}_{ij} = \delta_{ij} - k_{i} k_{j}/\vec{k}^2 \; , \nonumber \\ 
   \Sigma_{\mu \nu} = 2 \left(P^{(L)}\right)\indices{_{\left[ \mu \right.}^{\alpha}} \left(P^{(T)}\right)\indices{_{\left. \nu \right]}^{\beta}} \epsilon_{\alpha \beta \gamma} p^{\gamma} \; .
  \end{eqnarray}
The parity violating projector, $\Sigma$, corresponds to a contact term ambiguity in the definition of two point functions that must be specified to completely define the $(2+1)$-dimensional field theory of a conformal $U(1)$ current \cite{Witten:1995gf}.}

{\ Under the $S$-transformation the two point function has of course the same form, 
  \begin{eqnarray}
   \label{Eq:GenericTopologicalCorrelator}
      \left\langle J^{*}_{\mu}(p) J^{*}_{\nu}(-p) \right\rangle
  &=& \sqrt{p^2} \left[ C^{*}_{(\mathrm{T})} P^{(L)}_{\mu \nu} + C^{*}_{(\mathrm{L})} P^{(T)}_{\mu \nu} \right] + W^{*} \Sigma_{\mu \nu} 
  \end{eqnarray}
but with new scalar functions: $C^{*}_{(L)}, C^{*}_{(T)}, W^{*}$. These are given in terms of the original scalars by
   \begin{equation}
	 \label{Eq:Mapping}
	 C^{*}_{(L)} = \frac{{C}_{(L)}}{(2 \pi)^2 \left({C}_{(L)} {C}_{(T)} + {W}^2\right)} \; , \qquad
	 C^{*}_{(T)} = \frac{{C}_{(T)}}{(2 \pi)^2 \left({C}_{(L)} {C}_{(T)} + {W}^2\right)} \; , \; \nonumber
   \end{equation}
   \begin{equation}
	       W^{*} = -\frac{{W}}{(2 \pi)^2 \left({C}_{(L)} {C}_{(T)} + {W}^2\right)} \;, \; \; \; \; \;
  \end{equation}
Under a $T$-transformation however only the parity violating scalar is altered,
  \begin{equation}
    \label{Eq:TnTransformation}
    W^{*} = {W}+\frac{1}{2 \pi} \; ,
  \end{equation}
where we have normalised our currents suitably, leaving the other functions untouched.}

\subsection{Bulk formalism}

{\ The procedure of using holographic techniques to describe a theory with a dynamical gauge field based on the holographic description of a theory without a dynamical gauge field is called alternative quantisation. The reason is that the bulk description of the theories is the same but the quantisation procedure (i.e which fluctuations we quantise and which we treat as sources) changes.}

{\ The usual boundary condition (Dirichlet) imposed on the gauge field in AdS$_{4}$ stems from using a bulk action which after holographic renormalization and imposing the equation of motion has the property that
\begin{equation}
\delta \mathcal{S}_{\mathrm{D}}=\int _{\mathrm{boundary}} {J}^{\mu} \delta {A}_{\mu}
\end{equation}
where ${J}^{\mu}=\frac{\delta \mathcal{S}_{D}}{\delta {A}_{\mu}(\infty)}$ is interpreted as a conserved current in the CFT. Consistency of the variational principle (equivalently requiring no flux through the time-like boundary at infinity) requires imposing the condition that ${A}_{\mu}$ is fixed at the boundary. However this ``normal quantisation'' procedure is not the only consistent boundary condition one can impose on a gauge field in AdS$_{4}$. Both independent solutions of the equation of motion for the bulk gauge field are normalisable modes, and one can consider the theory with other boundary conditions (which again guarantee that no information is lost through time like infinity)\cite{Ishibashi:2004wx,Marolf:2006nd}.}

{\ Since $\partial^{\mu} {J}_{\mu}=0$ one can write $ {J}_{\mu}=\frac{1}{2\pi}\epsilon_{\mu \rho \nu}\partial^{\rho} {v}^{\nu}$, where ${v}_{\nu}$ is defined up to gauge transformations. Also it is convenient to define ${\mathcal{B}}_{\mu}\equiv\frac{1}{2\pi}\epsilon_{\mu \rho \nu}\partial^{\rho} {A}^{\nu}$. The most general boundary condition comes from the action 
  \begin{equation}
    \mathcal{S}_{\mathrm{generic}} = \mathcal{S}_{\mathrm{D}}+\frac{1}{2\pi}\int_{\mathrm{boundary}} \left[ a_{1} \epsilon_{\mu \rho \nu} {A}^{\mu}\partial^{\rho} {v}^{\nu} 
    +a_{2} \epsilon_{\mu \rho \nu}{A}^{\mu}\partial^{\rho} {A}^{\nu} + a_{3} \epsilon_{\mu \rho \nu} {v}^{\mu} \partial^{\rho}{v}^{\nu} \right] \; .
  \end{equation}
Now the variation of the action takes the form
  \begin{equation}
    \delta S_{\mathrm{generic}}=\int_{\mathrm{boundary}} (a_{s} {J}_{\mu}+b_{s} {\mathcal{B}}_{\mu})(c_{s}\delta {v}^{\mu}+d_{s} \delta {A}^{\mu})
  \end{equation}
where
  \begin{eqnarray}
    a_{s} d_{s}= 1 +a_{1} \; , \qquad b_{s} c_{s} = a_{1} \; , \qquad b_{s} d_{s} = 2 a_{2} \;, \qquad a_{s} c_{s} = 2 a_{3} \; .
  \end{eqnarray}
Evidently $a_{s} d_{s} - b_{s} c_{s} = 1$ and so form a $SL(2,\mathbb{R})$ matrix. The new boundary condition requires $c_{s} {v}^{\mu}+d_{s} {A}^{\mu}$ to be fixed, or in gauge invariant form
  \begin{eqnarray}
    \label{Eq:newcon}
    \mathcal{B}^{*}_{\mu}=c_{s} {J}_{\mu} + d_{s} {\mathcal{B}}_{\mu} = fixed, \; \;  \rightarrow \; \; \delta \mathcal{B}^{*}_{\mu}=0
  \end{eqnarray}
and the new current is just
  \begin{equation}
    \label{newj}
    J^{*}_{\mu}= a_{s} {J}_{\mu} + b_{s} {\mathcal{B}}_{\mu} \; .
  \end{equation}
The $S$-transformation is given by setting $a_{s} = d_{s} =0$ and $b_{s} = -c_{s}= 1$. The $T$-transformation has $a_{s}= b_{s}= d_{s}= 1$ and $c_{s}=0$. It is sometimes easier to take derivatives with respect to the gauge invariant combination \eqref{Eq:newcon}. The resulting objects are correlation functions of a gauge dependent quantity, but acting on it with an appropriate derivative operator gives the current correlation functions. Note that if the bulk theory has only integer electric and magnetic charges (which is what happens in string theory), for consistency  the $SL(2,\mathbb{R})$ transformations must be  restricted to a subset $SL(2,\mathbb{Z})$, acting on appropriately normalised quantities. }

{\ From equation \eqref{Eq:newcon} we see that the new current, $J^{*}_{\mu}$, is a current of particles carrying (with respect to the original definition) for each unit of the original charge, $c_s/d_s$ units of the original magnetic flux. This is precisely the realisation of anyons described in the introduction.}

{\ In this paper we explore theories which are connected to part of the phase space of the $D3/D5$ system through an $SL(2,\mathbb{Z})$ transformation. More specifically we look at this theory as the original theory at finite density and temperature and use a  $SL(2,\mathbb{Z})$ transformation to explore possible phenomena. For the pure $S$-transformation we prefer to view the transformed  theory as the original theory but coupled to a gauge field without a Maxwell term. We then holographically compute the field strength correlation function and interpret the quasi-normal modes as the spectrum of electromagnetic excitations (transverse) or plasmon excitations (longitudinal) in a finite temperature plasma of particles charged under a $U(1)$ gauge field and also strongly interacting via an $SU(N_{c})$ gauge field. For a more general $SL(2,\mathbb{Z})$ transformation we prefer to view the transformed theory as just another finite temperature CFT with a conserved charge carried by some excitations. In this case we 
holographically compute the current correlation functions and extract from them the collective excitations, conductivites etc. As explained 
above, after the $SL(2,\mathbb{Z})$ transformation, the charge carrying excitations are anyons. The new background charge density and magnetic field depend on the actual $SL(2,\mathbb{Z})$ transformation we have used.}

\section{Holographic model}

{\ We will use as our bulk theory the $D3/D5$ brane system where the $D5$-branes are probes of the background given by $N_{c}$ $D3$-branes at finite temperature. The contribution of a $D5$-brane to the bulk action is
  \begin{eqnarray}
   \label{Eq:Backgroundactiongeneral}
   \mathcal{S}^{(0)} &=& - T_{D5} \int d^{6} \xi \sqrt{-\det\left(g + F \right)} \;
  \end{eqnarray}
where $\xi$ are the embedding coordinates, $T_{D5}$ the tension of the $D5$ brane and $F$ the $U(1)$ world-volume field strength. We have absorbed a factor of $2\pi \alpha'$ into the field strength compared to the usual definition and thus it is dimensionless. As the $D5$ brane is treated as a probe we neglect its back-reaction upon the bulk metric which we must specify. We take the metric $g$ to be
  \begin{eqnarray}
   ds^2 & = & g_{tt}(r) dt^2  + g_{xx}(r) \left( dx^2+dy^2+dz^2\right) + g_{rr}(r) dr^2 + \ell^2 ds^2_{S^5} \; , \nonumber \\
	& = & -\frac{r^2}{\ell^2}f(r)dt^2+\frac{r^2}{\ell^2}\left(dx^2+dy^2+dz^2\right)+\frac{\ell^2}{r^2}\frac{dr^2}{f(r)} + \ell^2ds^2_{S^5} \; , 
	    \label{Eq:BackgroundMetric} \\
    f(r) & = & 1-\frac{r_H^4}{r^4} \; . \nonumber
  \end{eqnarray}
where $r=r_{H} = \pi T \ell^2$ with $\ell$ the AdS{} radius. We now choose $\ell \equiv 1$. The embedding we will consider is the usual massless black hole embedding with some background $U(1)_{b}$ charge, carried by bosonic and fermionic excitations, also considered recently in \cite{Brattan:2012nb}. The embedding is determined by the gauge field configuration since the scalar profiles are all trivial. It is well understood and we simply record the relevant results here, namely, the bulk $U(1)$ field strength is given by \cite{Kobayashi:2006sb,Karch:2007br},
  \begin{eqnarray}
    F = \mathrm{d} A = \frac{{d} \sqrt{g_{rr}|g_{tt}|}}{\sqrt{g_{xx}^2+d^2}} dr \wedge dt \; ,
    \label{Eq:BackgroundF}
  \end{eqnarray}
where ${d} \equiv \frac{\left\langle J^{t} \right\rangle}{\mathcal{N}_{5}}$, $\mathcal{N}_{5}=\frac{4\sqrt{\lambda}}{(2\pi)^3}N_{c}$,  and
  \begin{eqnarray}
   \left\langle J^{t} \right\rangle = \frac{\delta \mathcal{S}^{(0)}}{\delta A_t'} \; .
  \end{eqnarray}
This embedding has been proposed as the thermodynamically preferred state of the system for  all values of $d$ (see \cite{Chang:2012ek} for some controversy in the regard) and corresponds to the decoupling limit of the brane embedding displayed in table \ref{tab:branembedding}. The current displayed above does not have the correct length dimension to be a current due to our normalisation of $A$ in \eqref{Eq:Backgroundactiongeneral}. Subsequently the physical charge density is $\left(2 \pi \alpha'\right) d \mathcal{N}_{5}$.}

\begin{table}[t!]
  \centering
  \label{tab:branembedding}
  \begin{tabular}{|c||cccccccccc|} \hline
			 & $t$ & $x$ & $y$ & $z$ & $X^1$ & $X^2$ & $X^3$ & $X^4$ & $X^5$ & $X^6$\\ \hline \hline
    $N_c$ \,\,\, D$3$ & $\times$ & $\times$ & $\times$ & $\times$ & & &  &  & & \\
    $N_f$ \,\,\, D$5$ & $\times$ & $\times$ & $\times$ & & $\times$ & $\times$ & $\times$ & & & \\ \hline
  \end{tabular}
  \caption{The embeddings of the D3 and D5 branes in ten dimensional Minkowski space.}
\end{table}

{\ Now we turn to gauge fluctuations about the background field $A$ in \eqref{Eq:BackgroundF}. The boundary theory corresponding to \eqref{Eq:BackgroundMetric} and \eqref{Eq:BackgroundF} has explicit spatial $SO(2)$ rotation invariance and so all choices of the direction of spatial momentum for our fluctuation are equivalent. As such we shall turn on momentum in the $x$ direction, which perturbatively breaks this $SO(2)$, and then Fourier decompose our fluctuation $a$,
  \begin{eqnarray}
   a_{\mu}(r,x) &=& \int \frac{d\omega dk}{(2 \pi)^2} a_{\mu}(r,k) \exp \left( - i \omega t + i k x \right) \; ,
  \end{eqnarray}
where our transform conventions are as displayed. The quadratic action in terms of these fluctuations, $\mathcal{S}^{(2)}$, and the resulting equations of motion are straightforward to obtain but unilluminating, so we will omit them. We will record the equation for $a_r$, which in $a_r=0$ gauge is a constraint on the other bulk fluctuations,
  \begin{eqnarray}
    \label{eq:areom}
    \omega \, a_t' + u(r)^2 k \, a_x' =0 \; ,
  \end{eqnarray}
where
  \begin{eqnarray}
    u(r)^2 \equiv \frac{|g_{tt}|g_{rr} - A_t'^2}{g_{rr} g_{xx}} = \frac{|g_{tt}|g_{xx}}{g_{xx}^2+d^2} \; .
  \end{eqnarray}
With our choice of momentum, the gauge-invariant fluctuations are $a_y$ itself and also the bulk electric field (in Fourier space)~\cite{Kovtun:2005ev}
  \begin{eqnarray}
    E_{x}(r,\omega,k) \equiv k \, a_t(r,\omega,k) + \omega \, a_x(r,\omega,k) \; ,
  \end{eqnarray}
which is dual to the operator
  \begin{eqnarray}
    J_E \equiv k \, J^t + \omega \, J^x \; .
  \end{eqnarray}
In terms of these gauge-invariant fluctuations, using eq.~\eqref{eq:areom} and performing an integration-by-parts, we can write the quadratic action $\mathcal{S}^{(2)}$ as
  \begin{eqnarray}
    \label{eq:saquared}
    \mathcal{S}^{(2)} &=& \frac{\mathcal{N}_{5}}{2} \int dr \frac{d\omega dk}{(2 \pi)^2} \frac{|g_{tt}|}{u(r) g_{rr}^{1/2} g_{xx}^{1/2}} \times \\
		  & & \left[ \frac{1}{\omega^2 - u(r)^2 k^2} |E_{x}'|^2 - \frac{g_{rr}}{|g_{tt}|} |E_{x}|^2 - |a_y'|^2 + \frac{g_{rr}}{|g_{tt}|} \left( \omega^2 - u(r)^2 k^2 \right) |a_y|^2 \right] , \nonumber
  \end{eqnarray}
where $E_{x}$ and $a_y$ are generically complex. The equations of motion that follow from $\mathcal{S}^{(2)}$ are
  \begin{subequations}
    \begin{eqnarray}
    \label{eq:Eeom}
      E_{x}'' &+& \left [ \partial_r \log \left( \frac{|g_{tt}| g_{rr}^{-1/2}}{\left(\omega^2 - u(r)^2 k^2\right)u(r) g_{xx}^{1/2}} \right)\right] E_{x}' + \frac{g_{rr}}{|g_{tt}|}\left(\omega^2 - u(r)^2 k^2\right) E_{x} 
		   = 0 \; , \; \; \\
    \label{eq:ayeom}
      a_y'' &+& \left [ \partial_r \log \left(\frac{|g_{tt}|g_{rr}^{-1/2}}{u(r) g_{xx}^{1/2} }\right) \right] a_y'  + \frac{g_{rr}}{|g_{tt}|} \left(\omega^2 - u(r)^2 k^2\right)a_y =0  \; .
    \end{eqnarray}
  \end{subequations}
These match the equations of motion in refs.~\cite{Karch:2008fa,Myers:2008me,Davison:2011ek,Brattan:2012nb}. Notice in particular that the bulk equations of motion for $E_{x}$ and $a_y$ are decoupled and degenerate to the same equation when $k$ is taken to zero (i.e.~when the perturbation is homogeneous and thus respects $SO(2)$ spatial rotation invariance).}

\subsection{Alternative quantisation}

{\ Under a general $SL(2,\mathbb{Z})$ transformation the appropriate quasi-normal mode boundary condition\footnote{This is a slight abuse of standards as quasi-normal mode typically refers to the normal quantisation
condition.} can be written in terms of the bulk fields and a single parameter, which we will label as $n$, as
  \begin{equation}
    \label{Eq:mixedconditions}
    \lim_{r \rightarrow \infty} \left[ r^2 \delta F_{r\mu}-\frac{n}{2}\epsilon_{\mu \nu \rho}\delta F^{\nu \rho} \right] =0 \; .
  \end{equation}
Of course there is an appropriate $SL(2,\mathbb{Z})$ only for particular values of $n$ but we will compute as though it were a continuous parameter since the actual values corresponding to an $SL(2,\mathbb{Z})$ parameter depend on the precise value of $\mathcal{N}_{5}$. The spectrum of the quasi-normal modes, which are the poles of the retarded Green's function, are given by finding the normalisable modes relative to the boundary condition \eqref{Eq:mixedconditions} which only occur for particular pairs $(\omega,k)$. Thus the spectrum of the quasi normal modes depends only on the choice of $n$ and not on the precise $SL(2,\mathbb{Z})$ transformation. If we wish to compute the current two-point function however we need to completely specify the actual transformation.}

{\ Turning now to correlation function in order to compute the current-current correlator one needs to take derivatives of the on shell action with respect to the boundary conditions
  \begin{equation}
    \langle J^{*}_{\mu} J^{*}_{\nu} \rangle =\frac{\delta^2 S}{\delta(c_s {v}^{\mu}+d_s {A}^{\mu})\delta(c_s {v}^{\nu}+d_s {A}^{\nu})} \; .
  \end{equation}
In the treatment of the fluctuation analysis  this is actually computed by
  \begin{equation}
    \frac{\delta J^{*}_{\mu}}{\delta(c_s {v}^{\nu}+d_s {A}^{\nu})} \; .
  \end{equation}
However in the fluctuation analysis the boundary condition is given in terms of gauge invariant quantities $\delta \mathcal{B}^{*}_{\mu}=\delta(c_s {J}_{\mu}+d_s {\mathcal{B}}_{\mu})$.}

{\ Let $a^{*}$ be such that $\mathcal{B}^{*}_{\mu} = \frac{1}{2\pi} \epsilon_{\mu \nu \rho} \partial^{\nu} a^{\rho}_{*}$. Using the chain rule 
\begin{equation}
\frac{\delta J^{*}_{\mu}}{\delta \left(a^{*}\right)^{\nu}}=\frac{\delta J^{*}_{\mu}}{\delta \left(\mathcal{B}^{*}\right)^{\rho}}\frac{\delta \left(\mathcal{B}^{*}\right)^{\rho}}{\delta \left(a^{*}\right)^{\nu}}
\end{equation}
and that with the kinematics $p=(\omega,k,0)$, $\mathcal{B}^{*}_{x}$ and $\mathcal{B}^{*}_{t}$ are not independent, we get that 
\begin{equation}
  \langle J^{*}_{\mu}(p) J^{*}_{y}(-p) \rangle = -\frac{i\omega}{2\pi} \frac{\delta J^{*}_{\mu}(p)}{\delta \left(\mathcal{B}^{*}(p)\right)^{x}} \; , \ \
  \langle J^{*}_{\mu}(p) J^{*}_{x}(-p) \rangle = \frac{i\omega}{2\pi} \frac{\delta J^{*}_{\mu}(p)}{\delta \left(\mathcal{B}^{*}(p)\right)^{y}} \; ,\ \ \nonumber
\end{equation}
\begin{equation}
   \langle J^{*}_{\mu}(p) J^{*}_{t}(-p) \rangle = -\frac{ik}{2\pi}\frac{\delta J^{*}_{\mu}(p)}{\delta \left(\mathcal{B}^{*}(p)\right)^{y}} \; , \ \ \nonumber
\end{equation}
where we have defined $\epsilon_{txy} \equiv 1$. These equations allow us to compute the current two point function from the fluctuation analysis.}

{\ For low momentum and frequency it is often possible to solve \eqref{eq:Eeom} and \eqref{eq:ayeom} analytically. When this is not possible or unenlightening we shall resort to numerics. We now layout the numerical procedure \cite{Kaminski:2009dh} used to determine solutions to the bulk equations satisfying the mixed quantisation conditions\footnote{In this section we shall only display an explicit expression for the boundary source term and not the one-point functions. However when we reach section \ref{stntrans} we will show how the Green's function, and thus the one point functions, are obtained from our procedure.}. We shall employ the notation of \cite{Brattan:2012nb}. Eqs.~\eqref{eq:Eeom} and~\eqref{eq:ayeom} are second-order, hence for each field, $E_{x}$ and $a_y$, we need two boundary conditions to specify a solution completely. On the black hole horizon, a solution for $E_{x}$ or $a_y$ looks like a
linear combination of in-going and out-going waves, with some normalizations. The prescription for obtaining the \textit{retarded} Green's function requires that we choose our normalisations to remove any outgoing modes~\cite{Son:2002sd,Policastro:2002se,Skenderis:2008dh,vanRees:2009rw}. Let 
  \begin{eqnarray}
      \vec{V}(r,\omega,k)\equiv\begin{pmatrix} E_{x}(r,\omega,k) \\ a_y(r,\omega,k) \end{pmatrix}
  \end{eqnarray}
and at large $r$ identify
  \begin{eqnarray}
   \vec{V}(r,\omega,k) = \vec{V}^{(0)}(\omega,k) + \frac{1}{r} \vec{V}^{(1)}(\omega,k) + \mathcal{O}^{-2}(r) \; .
  \end{eqnarray}
For mixed quantisation a  boundary condition is given by fixing the combination \eqref{Eq:mixedconditions}. Using the relationship of \eqref{eq:areom} we find that (\ref{Eq:mixedconditions}) can be written in our current notation,
  \begin{eqnarray}
    \mathcal{N}_{5} \left[ \left( \begin{array}{cc} 1/p^2 & 0 \\ 0 & 1 \end{array} \right) \vec{V}^{(1)}(\omega,k) + i n \left( \begin{array}{cc} 0 & 1 \\ 1 & 0 \end{array} \right) \vec{V}^{(0)}(\omega,k) \right] = \vec{V}_{\mathrm{b}}=fixed
  \end{eqnarray}
to some value, denoted $\vec{V}_{\mathrm{b}}$, at the boundary. The first component of $\vec{V}_{\mathrm{b}}$ is equal to $\mathcal{B}^{*}_{x}$ and $\mathcal{B}^{*}_{t}$ when multiplied by $\omega$ and $-k$ respectively. For numerical purposes however it is preferable to fix all our boundary conditions at the future black hole horizon. The second boundary condition is then the normalisation of the ingoing wave at the horizon. The two ways to fix boundary conditions are related to each other by a change of basis transformation. The vector of near-horizon normalization factors, $\vec{V}_{\mathrm{nh}}$, is, when the temperature is non-zero,
  \begin{eqnarray}
    \label{eq:vnhdef}
    \vec{V}_{\mathrm{nh}} \equiv \lim_{r \rightarrow r_{H}} \, \exp \left( i \omega \int dr \sqrt{g_{rr}/|g_{tt}|} \right) \vec{V}(r,\omega,k).
  \end{eqnarray}
Notice that $\vec{V}_{\mathrm{nh}}$ is constant, independent of $r$, $\omega$, and $k$. On the right-hand-side of eq.~\eqref{eq:vnhdef}, the exponential factor is designed to \textit{cancel} the exponential factor that represents an
in-going wave at the future horizon.}

{\ We now pick two convenient values of $\vec{V}_{\mathrm{nh}}$ and solve the equations for each of these choices. This provides us with a basis of solutions in terms of which we can write any solution. The typical choices we have used in our numerics are: $\vec{V}^{(1)}_{\mathrm{nh}}=(1,0)^T$ and $\vec{V}^{(2)}_{\mathrm{nh}}=(0,1)^T$. Let us call the corresponding solutions $\vec{V}^{(1)}$ and $\vec{V}^{(2)}$ and use them to define a matrix $P(r,\omega,k)$ by
 \begin{eqnarray}
    \label{eq:pmatrixdef}
    P(r,\omega,k) \equiv \left(\vec{V}^{(1)}(r,\omega,k),\vec{V}^{(2)}(r,\omega,k)\right) \; .
 \end{eqnarray}
Using this matrix we can write any solution to the equations of motion with initial condition $\vec{V}_{\mathrm{nh}}$ at the horizon as
  \begin{eqnarray}
    \label{Eq:Diffproblem} 
    \vec{V}(r,p) = P(r,p) \, \vec{V}_{\mathrm{nh}}.
  \end{eqnarray}
In terms of the bulk to boundary propagator $P$ and the near horizon vector $\vec{V}_{\mathrm{nh}}$ we have
  \begin{eqnarray}
       \vec{V}_{\mathrm{b}}
   &=&  \mathcal{N}_{5} \lim_{r \rightarrow \infty} \left[ \left( \begin{array}{cc} 1/p^2 & 0 \\ 0 & 1 \end{array} \right) \left( -r^2 P'(r,\omega,k) \right) + i n \left( \begin{array}{cc} 0 & 1 \\ 1 & 0 \end{array} \right) P(r,\omega,k) \right]    
       \vec{V}_{\mathrm{nh}} \; . \nonumber
  \end{eqnarray}
The limiting case of $n=\infty$ is the normal quantisation condition \cite{Brattan:2012nb} while $n=0$ is the $S$-transform quantisation.} 

{\ We call any solution to the bulk equations with $\vec{V}_{\mathrm{b}}=0$ and complex frequency a quasi-normal mode. For a non-trivial solution with $\vec{V}_{\mathrm{b}}  \equiv 0$ and $\vec{V}_{\mathrm{nh}} \neq 0$ it must be the case that \cite{Kaminski:2009dh}
  \begin{eqnarray}
    \lim_{r \rightarrow \infty} \det \left[ \left( \begin{array}{cc} 0 & 1 \\ 1/p^2 & 0 \end{array} \right)  \left( -r^2 P'(r,\omega,k) \right) + i n P(r,\omega,k) \right] = 0
  \end{eqnarray}
which places a constraint on $\omega$ and $k$ yielding the dispersion relation of the mode\footnote{Note that the constraint for finding quasi-normal modes, as we have written it, is blind to poles at zero momentum and frequency.}. If we are interested in only normal or alternate boundary conditions we can take the limiting value of this expression. Setting $n \rightarrow \infty$ requires we set the determinant of $P$ to zero which is entirely standard. However, when we set $n=0$, it is very important to include the matrix prefactor to $P'$. It ensures that $d_{r} E_{x}$ is replaced by $d_{r} a_{x}$ or $d_{r} a_{t}$ when searching for the quasi-normal mode and cancels out an illusory light-like pole which would otherwise dominate the spectrum. Finally, we note that on setting $k=0$ the equations of motion degenerate into a single expression and thus finding the quasi-normal modes means solving a single equation equivalent to solving the boundary condition given by the upper or lower row of the above matrix.}

{\ We note that it will be useful throughout the remainder of the paper to set some conventions. A variable $x$ normalised by temperature will be denoted $\tilde{x}$. The object ${x}$ is evaluated in normal quantisation while, the quantity $x^{*}$ is calculated using $S$-transform quantisation. Finally if $x$ carries the sub- or superscript $n$, e.g.~$x_{n}$, it is determined by imposing mixed quantisation conditions.}

\section{A strongly coupled $(2+1)$-dimensional plasma}
\label{strans}

{\ In this section we consider the effects of gauging the external source vector field, $A^{(0)}$, in the absence of parity violation. We then employ this gauge field to probe properties of the finite density system. We compute various features from the two-point function of two $U(1)$ gauge operators such as the penetration length and the Debye mass, as well as the dispersion of the electromagnetic waves propagating in the plasma. }

{\ The currents are related to the field strength by $J_{\mu}=\frac{1}{2\pi}\epsilon_{\mu \nu \rho}\partial ^{\nu}A^{\rho}$. At zero temperature where $C_{L}=C_{T}=const$, one finds that for a wave vector in the $x$ direction, the only poles are lightlike and in the $F_{xy}$ and $F_{ty}$ correlators with no poles in the $F_{tx}$ correlator. This is what we expect in the vacuum for an electromagnetic wave. Given this identification, and also to agree with condensed matter literature, at non-zero temperature we shall refer to the longest lived quasi-particle excitations in the transverse electric field  correlator as the photon. The longest lived mode in the longitudinal electric field correlator at finite temperature will be called the plasmon.}

\subsection{The photon}

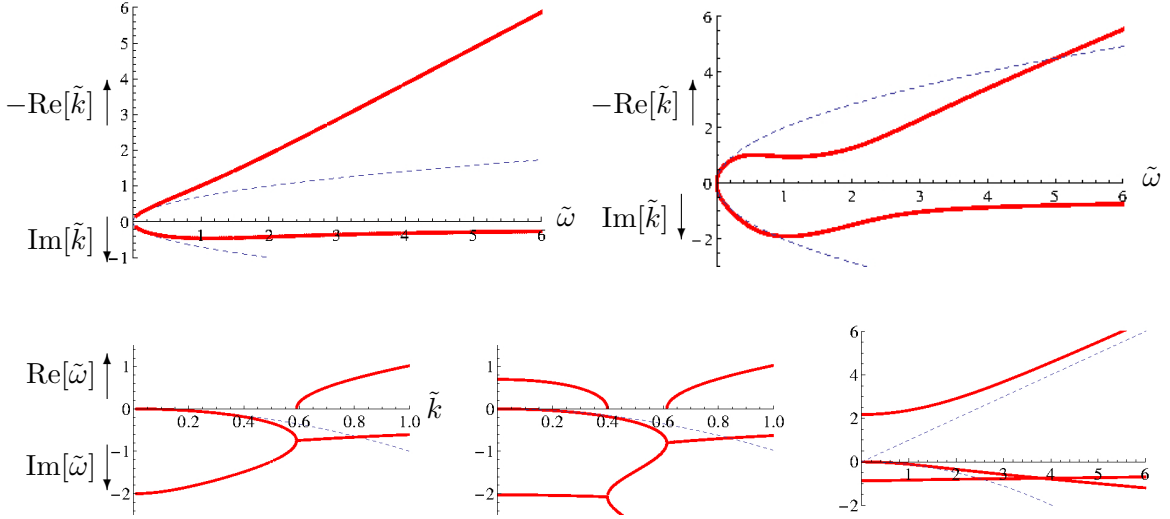
\begin{figure}
 \centering \hskip+3\unitlength
 \begin{subfigure}
  \centering
  \includegraphics[width=0.38\textwidth]{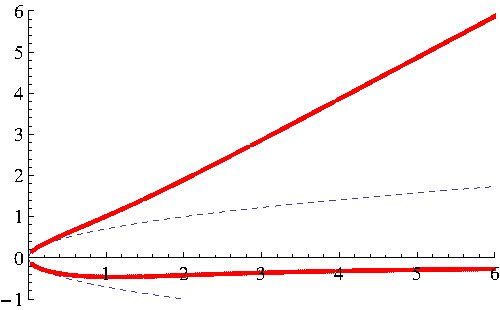}
 \end{subfigure} \hskip+11\unitlength
  \begin{subfigure}
  \centering
   \includegraphics[width=0.38\textwidth]{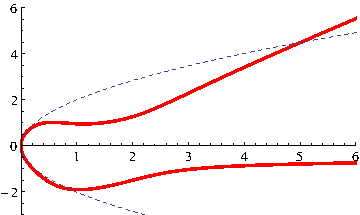}
 \end{subfigure} \\ \hskip+5\unitlength
 \begin{subfigure}
  \centering
  \includegraphics[width=0.27\textwidth]{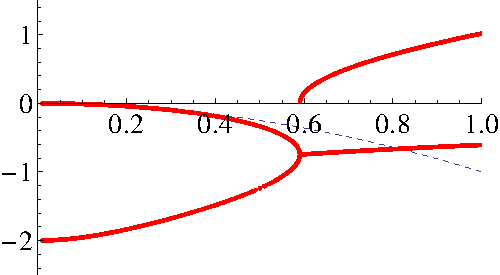}
 \end{subfigure} \hskip+3\unitlength
  \begin{subfigure}
  \centering
   \includegraphics[width=0.27\textwidth]{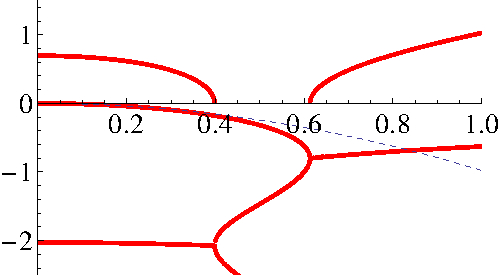}
 \end{subfigure} \hskip+3\unitlength
  \begin{subfigure}
  \centering
   \includegraphics[width=0.27\textwidth]{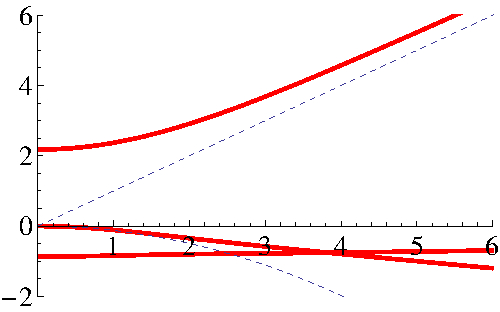}
 \end{subfigure} \\
 \begin{picture}(100,0)
  \put(2,41){\makebox(0,0){$-\Re[\tilde{k}]$}}
  \put(7,39){\vector(0,0){4}}
  \put(3,29){\makebox(0,0){$\Im[\tilde{k}]$}}
  \put(7,31){\vector(0,-4){4}}
  \put(53,41){\makebox(0,0){$-\Re[\tilde{k}]$}}
  \put(58,39){\vector(0,0){4}}
  \put(53,31){\makebox(0,0){$\Im[\tilde{k}]$}}
  \put(57,33){\vector(0,-4){4}}
  \put(47,31){\makebox(0,0){$\tilde{\omega}$}}
  \put(98,34.5){\makebox(0,0){$\tilde{\omega}$}}
  \put(3,17){\makebox(0,0){$\Re[\tilde{\omega}]$}}
  \put(7,15){\vector(0,0){4}}
  \put(3,9){\makebox(0,0){$\Im[\tilde{\omega}]$}}
  \put(7,11){\vector(0,-4){4}}
  \put(35.5,14.5){\makebox(0,0){$\tilde{k}$}}
 \end{picture}
 \vskip-2em
 \caption{The lowest lying quasi-normal modes of the transverse excitation for $\tilde{d}=0$ (top and bottom left), $\tilde{d}=2/10$ (bottom middle) and $\tilde{d}=5$ (top and bottom right). The upper diagrams are the momentum complexification of the Green's function at two choices of $\tilde{d}$ while the lower set of figures show the frequency complexification for three choices of $\tilde{d}$. For all figures we have chosen signs such that positive frequencies correspond to real parts of the dispersion relation. The red dots are numerical data while the blue dotted lines are the analytic expressions in \eqref{Eq:C}. In the bottom right figure we also include a blue line with gradient $1$ for comparison.} 
 \label{fig:dualdiffusion}
\end{figure}

{\ We shall begin by describing the lowest lying poles in the transverse electric field correlator,  organised by increasing $\tilde{d} = d/(\pi T)^2$. In fig.~\ref{fig:dualdiffusion} we display the lowest lying modes in this correlator at $\tilde{d}=0$, $\tilde{d}=2/10$ and $\tilde{d}=5$. We display both the case  where $\tilde{\omega} = \omega/(\pi T)$ is chosen to be real and the case where $\tilde{k} = k/(\pi T)$ is chosen to be real. The first choice represents the dispersion of electromagnetic waves in the plasma in response to an incident wave or due to time periodic phenomena. The second choice gives the excitations that arise when a spatially periodic phenomena occurs in the plasma.}

{\ Let us describe the behaviour of the dominant mode for real values of $k$. For all $\tilde{d}$ the hydrodynamical mode, the mode for which $\tilde{\omega}(\tilde{k} \rightarrow 0)\rightarrow 0$, is dominant for small $\tilde{k}$ or $\tilde{\omega}$. As $\tilde{k}$ is increased there is a transition to a mode with both real and imaginary parts for $\tilde{\omega}$, with $\Re (\tilde{\omega})\sim \tilde{k}$ and $\Im (\tilde{\omega}) \rightarrow 0$ for large $\tilde{k}$. This mode may look like a massive mode if $\tilde{d}$ is large enough. This can also be seen in the representaion using real $\tilde{\omega}$.}

\begin{figure}
 \centering \hskip-6\unitlength
 \begin{subfigure}
  \centering
  \includegraphics[width=0.43\textwidth]{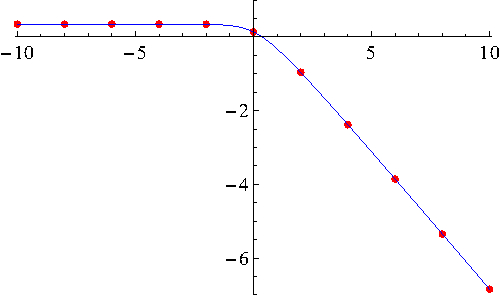}
 \end{subfigure} \hskip+5\unitlength
  \begin{subfigure}
  \centering
   \includegraphics[width=0.43\textwidth]{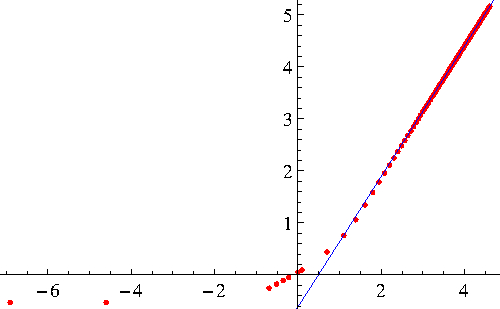}
 \end{subfigure}
 \begin{picture}(100,0)
  \put(23.5,33.5){\makebox(0,0){$\ln \left( \tilde{\omega}^{1/2} \tilde{\delta}_{p}(\omega) \right)$}}
  \put(76,34){\makebox(0,0){$\tilde{k}$}}
  \put(47.5,27){\makebox(0,0){$\ln(\tilde{d})$}}
  \put(97.5,8){\makebox(0,0){$\ln(\tilde{d})$}}
 \end{picture}
 \vskip-2em
 \caption{\textbf{Left:} The logarithm of the low frequency penetration depth normalised by the square-root of frequency against $\ln(\tilde{d})$ in a thermal strongly coupled $(2+1)$-dimensional field theory with background charge $\tilde{d}$. Red dots are numerical data while solid blue lines are analytic expressions. The blue line is obtained by solving \eqref{Eq:C} for $k(\omega)$ with $k$ complex and isolating the imaginary part of the dispersion relation. \textbf{Right:} The cross-over momentum against logarithm of the density. Red dots are numerical data while solid blue lines are analytic expressions.}
 \label{Fig:NonzerodLowMomentum1}
\end{figure}
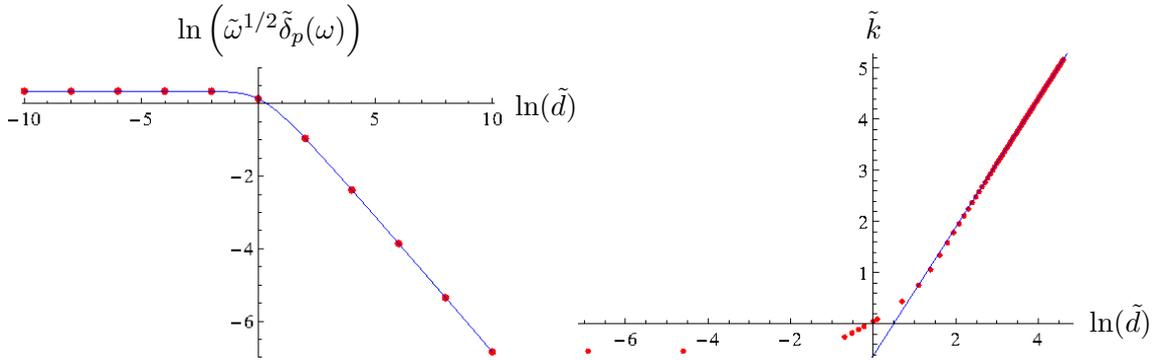

{\ The dispersion relation of the hydrodynamical mode can be solved for analytically in a small frequency and momentum analysis. This mode, for all values of $\tilde{d}$ is the dominant one for small $\tilde{\omega}$ and $\tilde{k}$. It satisfies the dispersion relation
  \begin{eqnarray}
   \label{Eq:C}
   \Gamma^{-1} \omega + i k^2 + \mathcal{O}(\omega^2,\omega k^2,k^4)= 0 \; , \qquad
   \Gamma =  \frac{1}{\sqrt{1+\tilde{d}^2}} \;_{2}F_{1}\left[\frac{1}{4},\frac{1}{2},\frac{5}{4};-\tilde{d}^2 \right] \; , \; \;
  \end{eqnarray}
for sufficiently small $\tilde{\omega}$ and $\tilde{k}$. An important physical quantity that can be obtained from this dispersion relation is the penetration length $\delta_{p}(\tilde{\omega})$ which measures how far the electromagnetic wave will penetrate the plasma. It is  defined as $1/\Im[\tilde{k}(\tilde{\omega})]$ where $\tilde{\omega}$ is chosen to be real. This is compared to numeric data in fig.~\ref{Fig:NonzerodLowMomentum1}. The analytic calculations are given in appendix \ref{diffusioncalc}. We note that at low frequency the penetration depth decreases as $\sim \tilde{d}^{-3/4}$ for $\tilde{d} \gtrsim 1$ and is roughly constant for $\tilde{d} \lesssim 1$. The penetration depth at arbitrary frequency, in other words outside the low frequency regime, can be obtained numerically as the imaginary parts of the real frequency dispersion relations. Examples for $\tilde{d}=0$ and $\tilde{d}=5$ are given in fig.~\ref{fig:dualdiffusion}.}

{\ As described above eventually the hydrodynamical mode ceases to be dominant and is overtaken by a mode which is approximately linear in $\tilde{k}$. We have tracked the crossing point in $\tilde{k}$ for $1/10 \leq \tilde{d} \leq 100$ as shown in fig.~\ref{Fig:NonzerodLowMomentum1}. We note at large $\tilde{d}$ the position of the crossover in $\tilde{k}$ grows as $\sim 0.54 \tilde{d}^{1.26}$. We do not expect the numbers displayed in this expression to have any element of universality and have simply recorded them for posterity. Indeed, we should expect them to depend strongly on the matter content of the bulk theory. However, the existence of the cross-over behaviour above should be more general and indeed has been seen often in the normally quantised system \cite{Brattan:2012nb}.}

\subsection{Plasma oscillations}

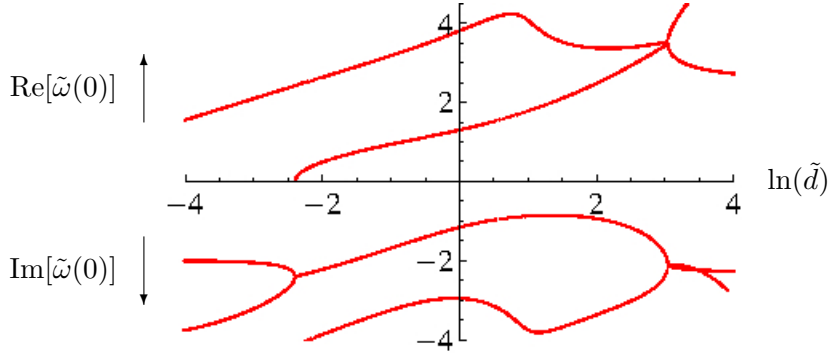
\begin{figure}[ht]
\center
\includegraphics[width=0.5\textwidth]{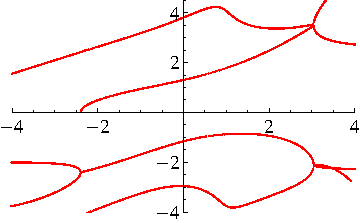}
\begin{picture}(100,0)
  \put(16,27){\makebox(0,0){$\Re[\tilde{\omega}(0)]$}}
  \put(23,24){\vector(0,0){6}}
  \put(16,11){\makebox(0,0){$\Im[\tilde{\omega}(0)]$}}
  \put(23,14){\vector(0,-4){6}}
  \put(80,19){\makebox(0,0){$\ln(\tilde{d})$}}
\end{picture}
\vskip-2em
\caption{The real (upper half) and imaginary (lower half) parts of the lowest quasi normal frequency $\tilde{\omega}(\tilde{k}=0)$ as a function of $\ln \tilde{d}$.}
\label{talk3}
\end{figure}

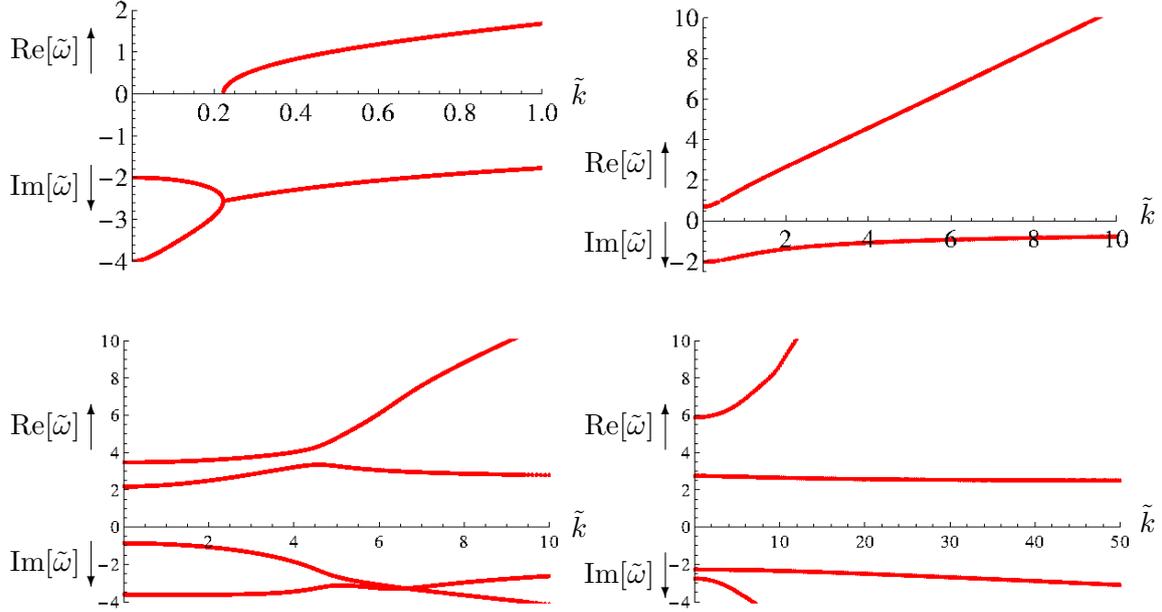
\begin{figure}
 \centering \hskip+5\unitlength
 \begin{subfigure}
  \centering
  \includegraphics[width=0.40\textwidth]{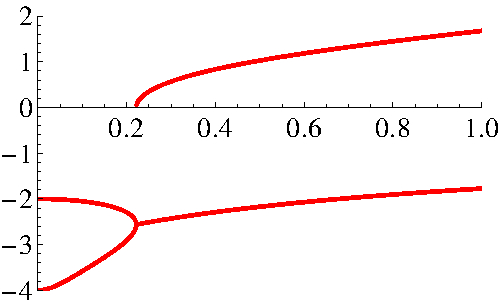}
 \end{subfigure} \hskip+8\unitlength
  \begin{subfigure}
  \centering
   \includegraphics[width=0.40\textwidth]{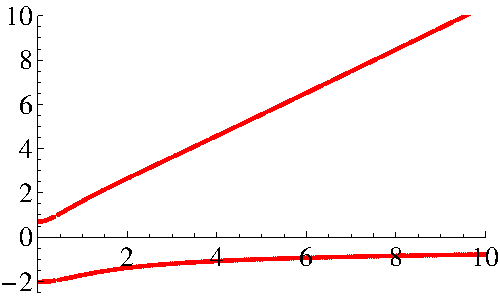}
 \end{subfigure} \\ \hskip+5\unitlength
  \begin{subfigure}
  \centering
   \includegraphics[width=0.40\textwidth]{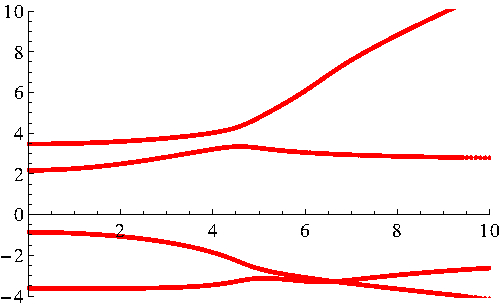}
 \end{subfigure} \hskip+8\unitlength
  \begin{subfigure}
  \centering
   \includegraphics[width=0.40\textwidth]{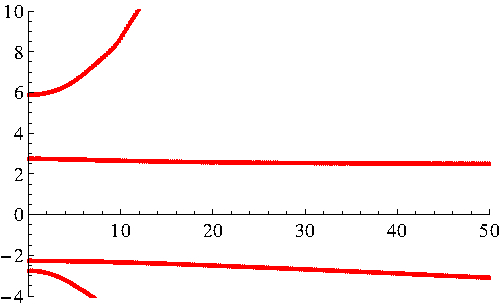}
 \end{subfigure}
 \begin{picture}(100,0)
  \put(3,54){\makebox(0,0){$\Re[\tilde{\omega}]$}}
  \put(7,52){\vector(0,0){4}}
  \put(3,42){\makebox(0,0){$\Im[\tilde{\omega}]$}}
  \put(7,44){\vector(0,-4){4}}
  \put(49.5,50.5){\makebox(0,0){$\tilde{k}$}}
  \put(3,21){\makebox(0,0){$\Re[\tilde{\omega}]$}}
  \put(7,19){\vector(0,0){4}}
  \put(3,9){\makebox(0,0){$\Im[\tilde{\omega}]$}}
  \put(7,11){\vector(0,-4){4}}
  \put(49.5,12.5){\makebox(0,0){$\tilde{k}$}}
  \put(53,44){\makebox(0,0){$\Re[\tilde{\omega}]$}}
  \put(57,42){\vector(0,4){4}}
  \put(53,37){\makebox(0,0){$\Im[\tilde{\omega}]$}}
  \put(57,39){\vector(0,-4){4}}
  \put(99,40){\makebox(0,0){$\tilde{k}$}}
  \put(53,21){\makebox(0,0){$\Re[\tilde{\omega}]$}}
  \put(57,19){\vector(0,4){4}}
  \put(53,8){\makebox(0,0){$\Im[\tilde{\omega}]$}}
  \put(57,10){\vector(0,-4){4}}
  \put(99,13){\makebox(0,0){$\tilde{k}$}}
 \end{picture}
 \vskip-2em
 \caption{The lowest lying quasi-normal modes of the longitudinal excitation for real $\tilde{k}$ and complex $\tilde{\omega}$. Displayed are $\tilde{d}=0$ (top  left)  , $\tilde{d}=2/10$ (top right), $\tilde{d}=5$ (bottom left) and  $\tilde{d}=50$ (bottom right).  The red dots are numerical data.}
 \label{fig:dualdiffusion2}
\end{figure}

{\ We now turn to the plasmon. This is an excitation which is a collective excitation of the plasma and thus vanishes at strictly zero temperature and zero density. The behavior of the leading excitation i.e.~that with the smallest imaginary part, changes as we change $\tilde{d}$. In figure~\ref{talk3} we give the quasi  normal mode at zero momentum behavior as we change $\tilde{d}$. As $\tilde{d}$
increases two purely imaginary modes (at $\tilde{k} =0$) come together and become a complex mode with a decreasing imaginary part and increasing real part as $\tilde{d}$ grows. At even larger $\tilde{d}$ two complex poles merge to become two other complex modes (at $\tilde{k} =0$) where the leading tends to having an almost constant real part as  $\tilde{d}$ grows.
In figure~\ref{fig:dualdiffusion2} we look at  what happens to these poles as one changes $\tilde{k}$. At small $\tilde{d}$ the purely imaginary mode comes together with another purely imaginary mode and becomes a complex mode with $\Re(\tilde{\omega}) \sim \tilde{k}$ at large $\tilde{k}$. At approximately $\ln(\tilde{d})=-2.5$ there is only a complex mode for  $\tilde{k} \geq 0$ which again has $\Re(\tilde{\omega}) \sim \tilde{k}$ at large $\tilde{k}$, and an increasing mass with increasing $\tilde{d}$. At an even larger $\tilde{d}$, for small $\tilde{k}$ the dominant mode has an almost constant $\Re(\tilde{\omega})$, but at larger $\tilde{k}$ the dominant mode switches to a mode with $\Re(\tilde{\omega}) \sim \tilde{k}$. As $\tilde{d}$ is increased the range of $\tilde{k}$ where the almost constant mode is the dominant one increases, but eventually at large enough $\tilde{k}$ another mode comes up from the complex plane to dominate which has $\Re(\tilde{\omega}) \sim \tilde{k}$.}

\subsubsection{Debye length}
{\ Now we turn to computing the Debye mass. This can be defined using the zero frequency electrostatic two point function. At zero frequency we find
  \begin{equation}
    \frac{1}{k^2} \langle E_{x}(0,k)E_{x}(0,-k) \rangle = \langle A_{0}(0,k)A_{0}(0,-k) \rangle,
  \end{equation}
and we note that the Fourier transform of this time component of the two point function will give the potential between two static point charges. Our system has conformal invariance at zero temperature and thus $\langle A_{0}(0,k)A_{0}(0,-k)\rangle \sim 1/\tilde{k}$ for $\tilde{k} \gg 1$. Hence the correlator begins as $\sim 1/r$ in vacuum but, as turning on temperature introduces a length scale, it becomes exponentially decaying at large $r$ for non-zero $T$. We call the $\tilde{d}$ dependent length scale in the exponential the ``Debye length''.  In summary at finite temperature the potential between static charges for our system will be
  \begin{eqnarray}
    V(r) &\sim& \left\{ \begin{array}{cc}
				\frac{1}{r} \; , & \frac{r}{r_{H}} \ll 1 \\
				\sqrt{\frac{1}{r}} \exp\left(-m_{\mathrm{D}} r\right) \; , & \frac{r}{r_{H}} \gg 1 
			     \end{array} \right. \; . \nonumber 
  \end{eqnarray}
}

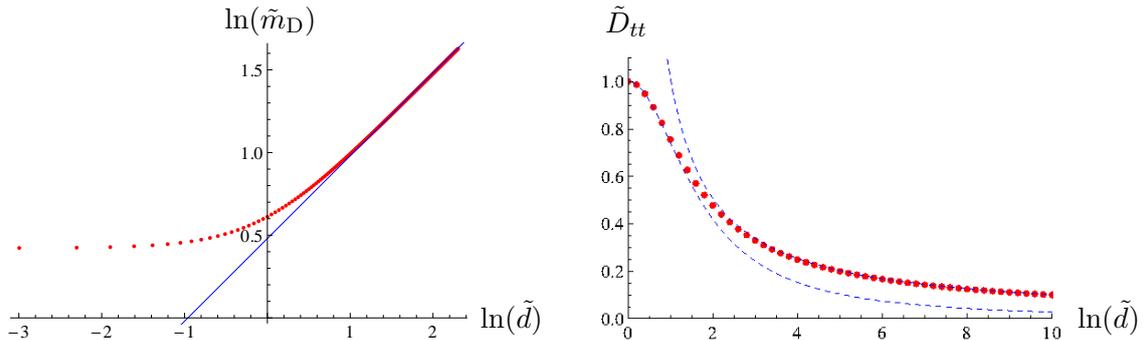
\begin{figure}
 \centering \hskip-6\unitlength
 \begin{subfigure}
  \centering
  \includegraphics[width=0.40\textwidth]{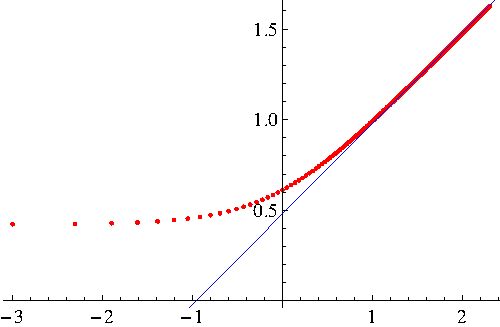}
 \end{subfigure} \hskip+10\unitlength
\begin{subfigure}
  \centering
   \includegraphics[width=0.4\textwidth]{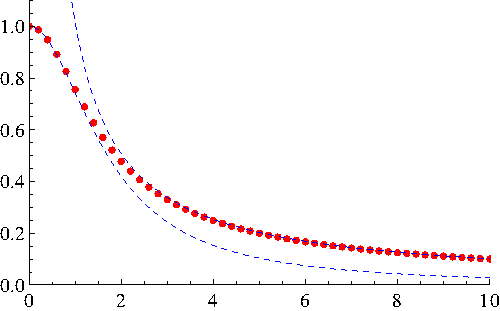}
 \end{subfigure}
 \begin{picture}(100,0)
  \put(45,7){\makebox(0,0){$\ln(\tilde{d})$}}
  \put(24,33){\makebox(0,0){$\ln(\tilde{m}_{\mathrm{D}})$}}
  \put(97,7){\makebox(0,0){$\ln(\tilde{d})$}}
  \put(55,33){\makebox(0,0){$\tilde{D}_{tt}$}}
 \end{picture} 
 \vskip-2em
 \caption{ \textbf{Left:} A plot of the logarithm of the Debye mass against the logarithm of charge density. The blue dots are numerical data while the solid red line is a fit to the large density behaviour. \textbf{Right:} The Fourier transform of the electrostatic potential as a function of $\tilde{k}$ at $\tilde{d}=0$. Dashed blue lines are fits at small $\tilde{k}$, $\sim\frac{1}{\tilde{k}^2+\tilde{m}^2}$ and at large $\tilde{k}$ $\sim 1/\tilde{k}$.}  
 \label{fig:masses}
\end{figure}

{\ The Debye length of our system is defined by $\frac{m_{\mathrm{D}}}{\pi T}=\tilde{m}_{\mathrm{D}} = -\Im[\tilde{k}(0)] $ where $\tilde{k}(\tilde{\omega})$ is the position of the lowest pole in the longitudinal correlator for real frequencies. This is displayed on the left of  fig.~\ref{fig:masses}. We have checked numerically that the pole is simple and paired (i.e.~there exists a second imaginary pole with opposite sign). On the right of  fig.~\ref{fig:masses} we also give the two point function of the zero component of the gauge field at zero frequency and real momenta.}
 
\section{Anyonic fluids}
\label{stntrans}

{\ We now turn to the more general boundary condition of \eqref{Eq:mixedconditions} labelled by $n$. The conserved charge is just the anyon number. The quasi-normal mode computation is insensitive to the actual $SL(2,\mathbb{Z})$ transformation since it only depends on $d_s/c_s=2\pi {\cal N}_5 n$. So the following results for the quasi-normal modes and their properties are true whenever $d_s/c_s$ has a particular value. However it is useful to have in mind a particular transformation when describing the results. So we will imagine that we are using the $SL(2,\mathbb{Z})$ transformation $T^L S T^K$, with ($K$ and $L$ are integer)
  \begin{eqnarray}
    a_s=-L,\ \  b_s=1 - KL \ \  c_s=-1, \ \ d_s=-K
  \end{eqnarray}
and so our parameter $n$ in \eqref{Eq:mixedconditions} takes the value $n=\frac{K}{2\pi {\cal N}_{5}}$. This means that the ground state of our system has $\langle \tilde{J}^{*}_{t} \rangle = \mathcal{N}_{5} {\tilde{d}} L$ and $\tilde{\mathcal{B}}^{*}_{t} = \mathcal{N}_{5} {\tilde{d}}$. As such our results are relevant for a thermal anyonic fluid at temperature $T$ and anyon density $\langle J^{*}_{t} \rangle$ in a magnetic field $\left(2\pi\mathcal{B}^{*}_{t}\right)$ whenever the filling fraction $\nu = \langle J^{*}_{t} \rangle/\left(2\pi\mathcal{B}^{*}_{t}\right)$ is equal to an integer $L$. In particular when $L=0$ this fluid consists of an equal number of anyons and anti-anyons in a background magnetic field.}

{\ As we change $n$ from zero we will get different finite temperature anyonic fluids at the same density in the same background magnetic field. However the fermionic and bosonic excitations of the original $D3/D5$ system are changed into anyonic excitations by a phase $\sim 1/n$ . We will treat $n$ as a continuous parameter but of course it can only take certain values. Thus the change of properties as we change $n$ indicates how the behaviour of the fluid depends on statistics of the theory's excitations. In order to avoid clutter we will label from now on $\tilde{b}^{*} \equiv \frac{\mathcal{B}^{*}_{t}}{{\cal N}_{5}}$.}

\subsection{Anyon correlator}

\begin{figure}
 \centering \hskip+5\unitlength
 \begin{subfigure}
  \centering
  \includegraphics[width=0.30\textwidth]{complexomegaofkd2_10B0_altEE}
 \end{subfigure} \hskip+1\unitlength
 \begin{subfigure}
  \centering
  \includegraphics[width=0.30\textwidth]{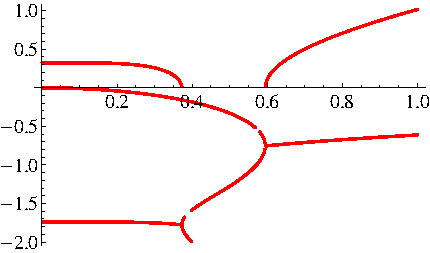}
 \end{subfigure} \hskip+1\unitlength
 \begin{subfigure}
  \centering
  \includegraphics[width=0.30\textwidth]{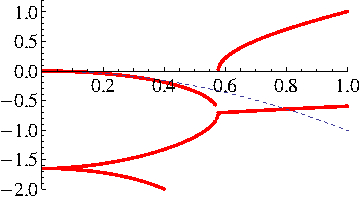}
 \end{subfigure} \\  \hskip+5\unitlength
 \begin{subfigure}
  \centering
  \includegraphics[width=0.30\textwidth]{complexomegaofkd5B0_altEE}
 \end{subfigure} \hskip+1\unitlength
 \begin{subfigure}
  \centering
  \includegraphics[width=0.30\textwidth]{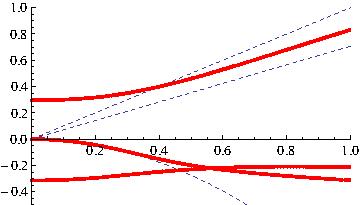}
 \end{subfigure} \hskip+1\unitlength
 \begin{subfigure}
  \centering
  \includegraphics[width=0.30\textwidth]{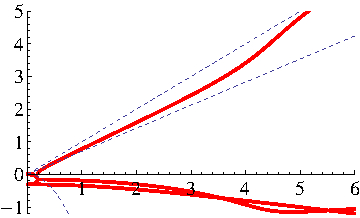}
 \end{subfigure}
 \begin{picture}(100,0)
  \put(3,44){\makebox(0,0){\small{$\Re[\tilde{\omega}]$}}}
  \put(3,39){\vector(0,4){3}}
  \put(3,33){\makebox(0,0){\small{$\Im[\tilde{\omega}]$}}}
  \put(3,38){\vector(0,-4){3}}
  \put(36,39){\makebox(0,0){\small{$\tilde{k}$}}}
 \end{picture}
 \vskip-2em
 \caption{Various plots of poles in the anyon current correlator. We have selected signs such that positive frequencies correspond to the real part of the dispersion relation for the displayed modes while negative frequencies give imaginary parts. Red dots represent numerical data while blue lines represent anayltic formulae. The top row has $\tilde{b}^{*}=2/10$ and $n=0,1,100$ (from left to right) while the bottom row has $\tilde{b}^{*}=5$ and $n=0,5, 100$ (again from left to right).
 In the bottom left there is a dashed blue line of gradient $1$, and in the bottom middle and right there are two lines of gradient $1$ and $1/\sqrt{2}$ respectively.}
 \label{fig:massive_zero_sound}
\end{figure}

{\ The pole structure of the correlators will be dependent on $\tilde{b}^{*}$ and $n$. Turning $n$ from small to large switches the pole structure from alternate quantisation ($n=0$) to normal quantisation ($n=\infty$) . The behaviour for various $\tilde{b}^{*}$ is displayed in fig.~\ref{fig:massive_zero_sound}. Not displayed in this figure is small $\tilde{b}^{*} \lesssim 1/10$ where there is little difference between $n=0$ and $n=\infty$. In this ultra low $\tilde{b}^{*}$ regime two purely imaginary modes come together at $\tilde{k} \sim 0.6$ and form a complex mode which asymptotes to having the light-like behaviour that governs the system at large $\tilde{k}$. As such we shall only further describe intermediate ($\tilde{b}^{*} \sim 1/10$) and high ($\tilde{b}^{*} \gg 1/10$) values of $\tilde{b}^{*}$.}

{\ At $\tilde{b}^{*} \sim 1/10$ and $n=0$ the behaviour of the lowest lying poles is slightly more complicated than the $\tilde{b}^{*} \lesssim 1/10$ case. There is a purely dissipative hydrodynamical mode, but instead of the purely imaginary mode below it, there is a complex mode starting at $\tilde{k}=0$. As $\tilde{k}$ increases it separates into two purely imaginary modes. One of these remains imaginary and sinks lower into the complex plane for the values of $\tilde{k}$ examined. The other combines with the hydrodynamical mode to become a complex mode which then asymptotes to a light-like pole. As $n$ is increased the splitting of the complex mode starts at smaller $\tilde{k}$ and as $n \rightarrow \infty$ the complex mode disappears and becomes two imaginary modes. All this is shown in the top of fig.~\ref{fig:massive_zero_sound}.}

{\ At larger $\tilde{b}^{*} \gtrsim 1/10$ the change in behaviour with $n$ is different. At $n=0$ the low-lying spectrum consists of a purely imaginary (diffusive) pole that descends deep into the complex frequency plane with increasing $\tilde{k}$ and a massive complex mode. For some finite $\tilde{k}$ this complex pole, which asymptotes to a light-like behaviour at large $\tilde{k}$, has smaller imaginary part than the diffusive pole and governs the late time behaviour.  As $n$ is increased a region of finite extent in $\tilde{k}$ appears where  the complex mode has a real part with gradient $\approx \pm 1/\sqrt{2}$. This looks like a massive zero sound mode for the anyons. At larger $\tilde{k}$ there is a kink and the dispersion then becomes light-like. The range of $\tilde{k}$ for which the zero sound mode behaviour is seen increases with increasing $n$ and $\tilde{b}^{*}$. For any given $\tilde{b}^{*}$ there is an $n$ small enough where this behaviour is absent and similarly for a given $n>0$ there 
is a small 
enough $\tilde{b}^{*}$ where no zero sound behaviour is seen, but rather just a lightlike pole. For a fixed $\tilde{b}^{*}$ large enough for zero sound to exist as we increase $n$ the mass of the zero sound mode decreases. For larger $n$ at small enough $\tilde{k}$ the lower mode is still complex. However at some small value of $\tilde{k}$ it splits into two imaginary modes, one of which sinks deeper into the complex plane while the other joins up with the hydrodynamical mode as $\tilde{k}$ increases. All this is seen in the bottom row of fig.~\ref{fig:massive_zero_sound}.}

{\ As a final comment on the pole structure we note that the emergence of the massive pole at non-zero $\tilde{b}^{*}$, fig.~\ref{fig:dualdiffusion}, is similar to what is seen in the normally quantised $D3/D5$ and $D3/D7'$ systems at finite density and magnetic field \cite{Brattan:2012nb,Jokela:2012vn}. For vanishing magnetic field in that system the collisionless mode (whether it be zero sound at large $\tilde{d}$ or light-like at small $\tilde{d}$) has a massless dispersion relation after the cross-over. When the magnetic field is non-zero it acquires a small mass but the system has sufficient thermal energy to excite this mode regardless so we still see the purely imaginary diffusion pole connect up with a pole from deeper in the complex plane to become complex. As magnetic field increases further the mass of the mode becomes too large to be overcome by thermal effects and the diffusion poles ceases to connect with another pole and simply sinks lower into the complex plane as $\tilde{k}$ increases. Now 
at 
some 
value of $\tilde{k}$ the 
collisionless pole has a smaller imaginary part than the diffusion pole and instead governs the late time behaviour of the system. See \cite{Brattan:2012nb,Jokela:2012vn} for further discussion.}

\subsubsection{Diffusion constant}

{\ A feature that is common to perturbations of interacting thermal systems by conserved current operators is the existence of a diffusion regime at sufficiently long-times and low momenta. This regime is governed by the behaviour of poles close to the origin.}

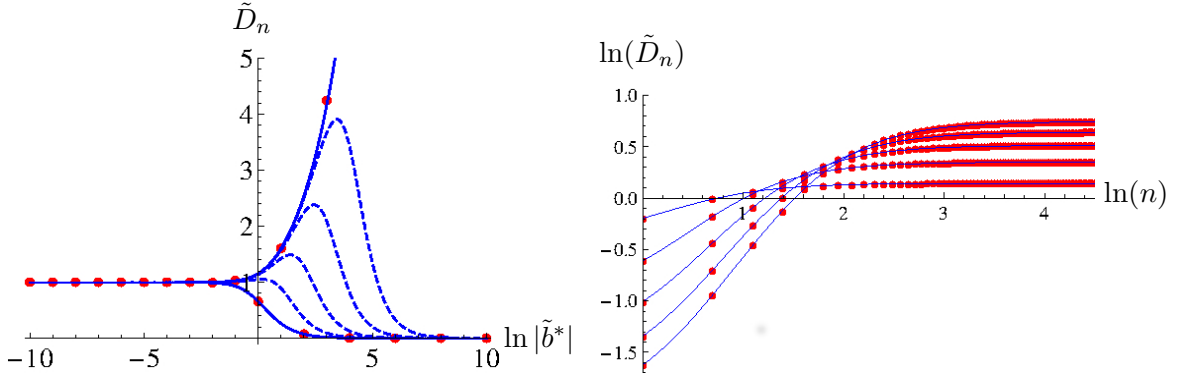
\begin{figure}
 \centering \hskip-6\unitlength
 \begin{subfigure}
  \centering
  \includegraphics[width=0.43\textwidth]{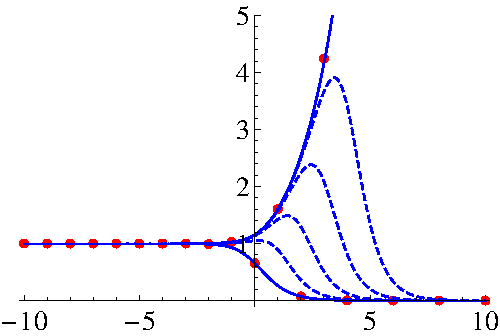}
 \end{subfigure} \hskip+7\unitlength
  \begin{subfigure}
  \centering
    \includegraphics[width=0.43\textwidth]{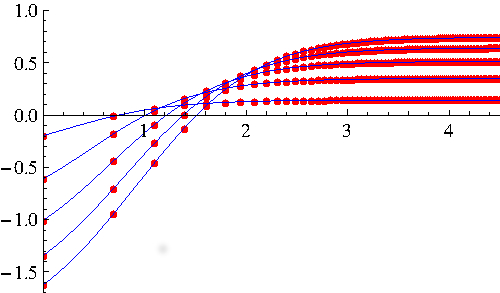}
 \end{subfigure}
 \begin{picture}(100,0)
  \put(21,36){\makebox(0,0){$\tilde{D}_{n}$}}
  \put(46,8){\makebox(0,0){$\ln |\tilde{b}^{*}| $}}
  \put(55,33){\makebox(0,0){$\ln(\tilde{D}_{n})$}}
  \put(98,20.5){\makebox(0,0){$\ln(n)$}}
 \end{picture}
 \vskip-2em
 \caption{\textbf{Left:} The diffusion constant of the anyonic current at various $n$ against $\ln |\tilde{b}^{*}|$.  The lines have increasing $n$ from bottom to top. Red dots represent numerical data while blue lines are from analytic formulae. The uppermost solid blue line  is the diffusion constant as $n \rightarrow \infty$ and the lowermost solid blue line is the $n=0$ case. \textbf{Right:} The diffusion constants for the anyonic current  against $\ln(n)$. Starting on the vertical axis and proceeding from the least negative curve to the most negative they have values of $\tilde{b}^{*}=1,2,3,4,5$. The analytical formulae for the solid blue lines are given in \eqref{Eq:diffusionvariousn}.}
 \label{Fig:NonzerodDiffusionConstantsvariousn}
\end{figure}

{\ The diffusion constant of the anyon current as a function of  $n$ is computed in appendix \ref{diffusioncalc}, and the result is
  \begin{eqnarray} 
    \label{Eq:diffusionvariousn}
    \tilde{D}_{n} =     \frac{\left(\tilde{b}_{*}^2+1\right) \left(n^2+2\right) \;_{2}F_{1}\left[-\frac{1}{4},1,\frac{1}{4};-\tilde{b}_{*}^2\right]- \left(2 \tilde{b}_{*}^2+n^2+2 \right)}{2 \tilde{b}_{*}^2 \left(\tilde{b}_{*}^2+n^2+1\right)}     \; . 
  \end{eqnarray}
Note that for a given $\tilde{b}^{*}$ it has a maximum value at $n=0$ and thus anyons are much less efficient at depositing charge into the ground state. Here we have an infinite class of diffusion constants for differing types of anyonic excitations. The analytic expression for the diffusion constant at various choices of $n$ against numerical data is depicted in fig.~\ref{Fig:NonzerodDiffusionConstantsvariousn}. Notice also that the above behaviour with $n$ makes the normal quantisation diffusion something of a denegerate case. For sufficiently large $\tilde{b}^{*}$, except when $n=\infty$, the diffusion constant tends to the $S$-transform diffusion constant.}

\subsubsection{Conductivities}

{\ At  zero temperature, density and magnetic field the current correlator obtained from gauging the external source when the original theory contains a non-zero Chern-Simon's number takes the form
  \begin{eqnarray}
   \left\langle J^{*}_{\mu}(p) J^{*}_{\nu}(-p) \right\rangle
   = \frac{1}{(2 \pi)^2 \mathcal{N}_{5}} \left[ \sqrt{p^2} \left(\frac{1}{1+n^2}\right) P_{\mu \nu} + \left(\frac{n}{1+n^2} \right) \Sigma_{\mu \nu} \right] \; .
  \end{eqnarray}
We see that in the large $n$ limit the correlator is vanishing unless we rescale with $n^2$. This scaling reproduces the two point function of the currents in normal quantisation with Chern-Simon's level $n$. In the numerics, at large $n$, $\tilde{\omega}$ and $\tilde{k}$, we expect our anyon correlators to have vanishing residues but a non-trivial pole structure. In particular, the large frequency AC conductivities should tend to zero for increasing $n$.}

{\ To extract the AC conductivities of our theory we will need to numerically compute the Green's function. The two point function is given in terms of the matrix $P(r,p)$ by
  \begin{eqnarray}	
     G_{n}(p) &=& \frac{1}{(2 \pi)^2 \mathcal{N}_{5}} \lim_{r \rightarrow \infty} \left\{ \left( \begin{array}{cc} 0 & -\omega \\ 1 & 0 \end{array} \right) P(r,p) \times \right. \nonumber \\
	      &\;& \left. \left[ \left( \begin{array}{cc} 1/p^2 & 0 \\ 0 & 1 \end{array} \right) \left( \frac{r^2}{\ell} P'(r,p) \right) + i n \left( \begin{array}{cc} 0 & 1 \\ 1 & 0 \end{array} \right) P(r,p) \right]^{-1}     
		    \left( \begin{array}{cc} 0 & 1 \\ \omega & 0 \end{array} \right) \right\} \nonumber \\
	      &=& \left( \begin{array}{cc}
			             \langle J^{*}_{x}(p) J^{*}_{x}(-p) \rangle  & \langle J^{*}_{x}(p) J^{*}_{y}(-p) \rangle  \\
			             \langle J^{*}_{y}(p) J^{*}_{x}(-p) \rangle  & \langle J^{*}_{y}(p) J^{*}_{y}(-p) \rangle  \\
			            \end{array} \right) \; .
  \end{eqnarray}
If the reader would prefer components with $t$ as opposed to $x$ they need only replace the explicit factors of $\omega$ by $-k$. The AC conductivities are related to the mixed Green function $G_{n}$ by
  \begin{eqnarray}
    \sigma_{n}(\omega) = \frac{1}{i\omega} G_{n}(\omega,\vec{0}) = \left( \begin{array}{cc} 
											    \sigma^{(L)}_{n}(\omega) & \sigma^{(H)}_{n}(\omega) \\
											    -\sigma^{(H)}_{n}(\omega) & \sigma^{(L)}_{n}(\omega) \\
                                                                                          \end{array} \right) \; 
  \end{eqnarray}
where the transverse and longitudinal conductivities are equal due to rotation invariance. The DC conductivity is given by the $\omega \rightarrow 0$ limit of the above expression.}

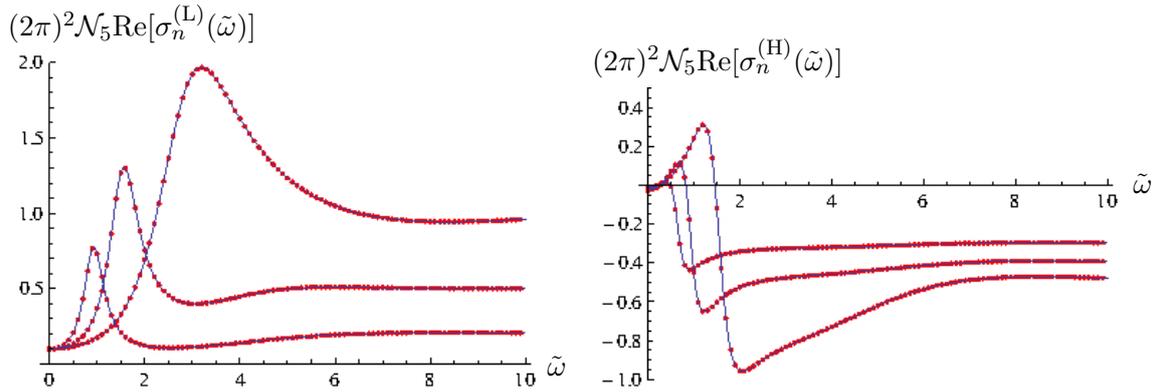
\begin{figure}
 \centering \hskip-2\unitlength
 \begin{subfigure}
  \centering
  \includegraphics[width=0.45\textwidth]{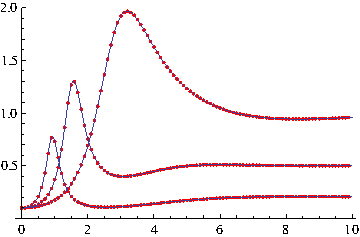}
 \end{subfigure} \hskip+4\unitlength
  \begin{subfigure}
  \centering
    \includegraphics[width=0.45\textwidth]{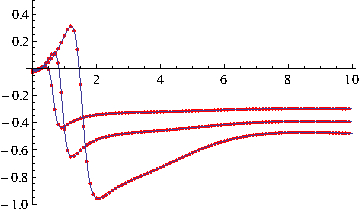}
 \end{subfigure}
 \begin{picture}(100,0)
  \put(11,37){\makebox(0,0){$(2 \pi)^2 \mathcal{N}_{5} \Re[\sigma^{\mathrm{(L)}}_{n}(\tilde{\omega})]$}}
  \put(48,7){\makebox(0,0){$\tilde{\omega}$}}
  \put(62,34){\makebox(0,0){$(2 \pi)^2 \mathcal{N}_{5} \Re[\sigma^{\mathrm{(H)}}_{n}(\tilde{\omega})]$}}
  \put(99,23){\makebox(0,0){$\tilde{\omega}$}}
 \end{picture}
 \vskip-2em
 \caption{Figures comparing analytic expressions for the longitudinal (and transverse) and Hall conductivities for the anyonic fluid at background magnetic field $\tilde{b}^{*}$. Red dots represent numerical data while the solid blue lines are analytical formulae. \textbf{Left:} The real part of the longitudinal AC conductivity at $\tilde{b}^{*}=10$ against $\tilde{\omega}$ for three values of $n$. The plot with the largest 
range has $n=0$, that with the smallest $n=2$ and the intermediate line has $n=1$. \textbf{Right:} The real part of the AC Hall conductivity at $\tilde{b}^{*}=10$ against $\tilde{\omega}$ for three values of $n$. The plot with the largest range has $n=1$, that with the smallest $n=3$ and the intermediate line has $n=2$. The $n=0$ line would lie along the horizontal axis.}
 \label{Fig:NonzerodConductivitiesvariousn}
\end{figure}

{\ As a cross-check of the numeric results obtained from this expression, given the transformation formulae of appendix \ref{diffusioncalc}, we can obtain the conductivities of the $T^{L}ST^{K}$-transformed system from the original system. The blue ``analytic'' lines of fig.~\ref{Fig:NonzerodConductivitiesvariousn} are obtained via this transformation of the original system AC conductivities. The DC conductivities of the transformed system can be obtained analytically and they are
  \begin{eqnarray}
   \label{Eq:STnconductivities}
   (2 \pi)^2 \sigma^{(\mathrm{L})}_{n} = \frac{1}{\mathcal{N}_{5}} \left(\frac{\sqrt{1+ \tilde{b}_{*}^2}}{1 + n^2 + \tilde{b}_{*}^2}\right) \; , \qquad 
   (2 \pi)^2 \sigma^{(\mathrm{H})}_{n} = \frac{1}{\mathcal{N}_{5}} \left(\frac{n}{1 + n^2 + \tilde{b}_{*}^2}\right) + (2 \pi) L\; .
  \end{eqnarray}
The DC and AC conductivities of the original system are outlined in appendix \ref{normalquantisation}.}

{\ Note that the anyonic fluid  system has a $\tilde{b}^{*}$ dependent Hall conductivity, which does not vanish at $\tilde{b}^{*}=0$. Multiplying the DC conductivities by $n^2$ and taking $n \gg \sqrt{1+\tilde{b}_{*}^2}$ the longitudinal anyon conductivity of \eqref{Eq:STnconductivities} takes the $n=0$ value. Similarly the Hall conductivity tends to $n$ when $L=0$. The necessity of the $n^2$ prefactor in obtaining the normal quantisation conductivities is a manifestation of the fact that the Green's function of the $ST^{K}$ system are not equal to that of the normally quantised system in the large $n$ limit despite having exactly the same pole structure. This difference is due to a difference in residues.}

{\ In fig.~\ref{Fig:NonzerodConductivitiesvariousn} we  display the real parts of the frequency dependent generalisation of these expressions for $L=0$ i.e.~the zero anyon density fluid. The finite $L$ system is given by shifting the horizontal axis of the Hall conductivity vertically by an integer times by $2 \pi \mathcal{N}_{5}$. The imaginary parts of the conductivities can be recovered via the Kramers-Kronig relations. We see that as we increase $n$ for a fixed $\tilde{b}^{*}$ the peak in the longitudinal conductivity, which should be thought as the shifted Drude peak, shrinks and moves closer to small $\tilde{\omega}$. If instead we fix $n$ and increase $\tilde{b}^{*}$ we find that the location of the peak moves towards large $\tilde{\omega}$ roughly like $\tilde{b}_{*}^{1/3}$. At zero $\tilde{b}^{*}$ the AC conductivities are independent of the frequency.}

{\ Another interesting feature of these expressions is that, for a given $\tilde{b}^{*}$, the longitudinal conductivity is always maximised by $n=0$ while the Hall conductivity is maximized at some non-zero value of $n$. Consider any $O(2)$ invariant $(2+1)$-dimensional field theory with matter fields coupled to an external gauge field. Regardless of the matter content, when we perform an $ST^{K}$ transformation we will find there exists an $n_{\mathrm{crit.}}$ where the Hall conductivity is maximized for fixed values of the background parameters. We can even determine its value to be $n_{\mathrm{crit.}} = \pm \sigma_{(L)}$. Moreover at this value of $n$ the longitudinal and Hall conductivities will be equal. For $n>n_{\mathrm{crit.}}$ the Hall conductivity is larger than the longitudinal conductivity and vice versa for $n<n_{\mathrm{crit.}}$. In our case $n_{\mathrm{crit.}} = \pm \sqrt{1+\tilde{b}_{*}^2}$.}

\section{Discussion}
\label{discussion}

{\ We have considered the alternative quantisation of the $D3/D5$ system and used it to explore properties of a strongly coupled charged plasma and strongly coupled anyonic fluids. The $S$-transform of the $D3/D5$ system was used as a model for charged matter interacting with a $U(1)$ gauge field in the large coupling regime. We computed the dispersion relationship of the propagating electromagnetic modes as the density and temperature are changed and found a mode at large densities with momentum independent behaviour for a large range of momentum. This mode we identified as the plasmon. Additionally we calculated the Debye length in this model.}

{\ We then considered a more general $SL(2,\mathbb{Z})$ transformation yielding a strongly interacting anyonic fluid. We studied its transport properties as functions of the statistics of the anyons and the background magnetic field. In particular we have demonstrated how to calculate the diffusion constant for any anyonic fluid obtained by an $SL(2,\mathbb{Z})$ transform given the diffusion constant of the original system. At large magnetic field we found modes in the spectrum which gain a mass that are qualitatively similar to those seen in the normally quantised $D$3/$D$5 system at finite density for sufficiently large magnetic field.}

{\ One of the more interesting outcomes from our investigation is the fate of zero sound as the statistics of our fluid changes. For sufficiently large $\tilde{b}$ and $n$ we found a zero sound mode which is the analogue of zero sound in \cite{Karch:2008fa}. As we tune $n$ down the statistics of our theory changes from that of fermions and bosons to anyons. This results in the zero sound region disappearing from the excitation spectrum of our theory. It would also be interesting to understand this effect at weak coupling.} 

{\ There are several avenues for future study that could provide interesting insight into the nature of anyons at strong coupling. For example it would be interesting to understand the nature of the anyonic fluid at zero magnetic field. To do this in our model requires that we generalise our embedding to include a background magnetic field. Such backgrounds \cite{Evans:2010hi,Pal:2010gj,Jensen:2010vx,Jensen:2010ga}, and their normal quantisation perturbations \cite{Goykhman:2012vy,Brattan:2012nb}, are well understoood. This generalisation to our results will follow soon.}

\acknowledgments

{\ DB is supported by in part by the Israeli Science Foundation (ISF) grants 2016577 and 392/09, University of Haifa President's Postdoctoral Fellowship and a Fine Fellowship. The work of GL was supported in part by the ISF under grant no.~392/09 and in part by a grant from the GIF, the German-Israeli Foundation for Scientific Research and Development under grant no.~1156-124.7/2011. DB and GL would like to thank Irene Amado, Oren Bergman, Kristan Jensen, Niko Jokela, Matthew Lippert, Caterina Riconda and Amos Yarom for useful discussions.}

\appendix

\section{Diffusion constants calculation}
\label{diffusioncalc}


{\ In the introduction we considered the decomposition of a two point function of currents into particular tensor structures $P_{(\mathrm{L})}$, $P_{(\mathrm{T})}$ and $\Sigma$. Subsequently we
reviewed how normal and alternative quantisation of a bulk gauge field relate the decomposition of the current-current correlator to two point function of field strengths. This perscription \eqref{Eq:Mapping} is completely generic as it relies only upon symmetries of the theory. However, it is more interesting to understand how physical parameters like the conductivities map onto eachother. In this appendix we shall determine the mappings between physical observables such as the conductivity tensor and hydrodynamic\footnote{Strictly the formalism we consider is not hydrodynamics
but simply a long wavelength and low frequency expansion as the total (background+brane) stress tensor does not participate in the leading dynamics of the probe brane due to taking the  probe limit. Additionally we only consider  solving the linearised charge conservation equation while fluid dynamics proper is the study of non-linear solutions to this equation.} transport coefficients.}

{\ We begin by isolating the DC and AC conductivities. Taking the spatial momentum to zero our decomposition of the normal quantisation two point function implies
  \begin{eqnarray}
    \left\langle {J}_{i}(\omega,\vec{0}) {J}_{j}(-\omega,\vec{0}) \right\rangle = \pm i \omega \left[ {C}_{(\mathrm{L})}(\omega,\vec{0}) \delta_{ij} + {W}(\omega,\vec{0}) \epsilon_{ij} \right] \; , \nonumber \\ \left\langle {J}_{t}(\omega,\vec{0}) {J}_{i}(-\omega,\vec{0}) \right\rangle = \left\langle {J}_{i}(\omega,\vec{0}) {J}_{t}(-\omega,\vec{0}) \right\rangle = 0 \;,
    \qquad \left\langle {J}_{t}(\omega,\vec{0}) {J}_{t}(-\omega,\vec{0}) \right\rangle = 0 \; ,
  \end{eqnarray}
where $i,j$ are spatial indices and $\omega \epsilon_{ij} : = \epsilon_{\mu \nu \rho} p^{\rho}$. The AC conductivities are then defined by
  \begin{eqnarray}
    \sigma_{ij}(\omega) = \frac{1}{i \omega} \left\langle {J}_{i}(\omega) {J}_{j}(-\omega) \right\rangle 
     = \left[  {C}_{(\mathrm{L})}(\omega,\vec{0}) \delta_{ij} + {W}(\omega,\vec{0}) \epsilon_{ij} \right] \; , \nonumber \\
    {\sigma}_{(\mathrm{L})} = {\sigma}_{(\mathrm{T})} = {C}_{(\mathrm{L})}(\omega,\vec{0}) \; , \qquad \sigma_{(\mathrm{H})}(\omega) = {W}(\omega,\vec{0}) \; , 
  \end{eqnarray}
with the DC conductivities being given by the zero frequency limit of these expressions. DC conducitivities will be denoted by dropping the frequency dependence e.g. $\sigma_{(\mathrm{L})}$ is the DC longitudinal conductivity while $\sigma_{(\mathrm{L})}(\omega)$ is the AC longitudinal conductivity. For the $S$-transform of the normal quantisation system the decomposition into longitudinal, transverse and Hall conductivities is identical. The conductivities in the $S$-transformed system are related to those of the original system by
  \begin{eqnarray}
   \label{Eq:ACconductivities}
   \sigma^{*}_{(\mathrm{L})}(\omega) = \frac{1}{(2\pi)^2} \frac{{\sigma}_{(\mathrm{L})}(\omega)}{{\sigma}_{(\mathrm{L})}^{2}(\omega) + {\sigma}_{(\mathrm{H})}^2(\omega)} \; , \qquad
   \sigma^{*}_{(\mathrm{H})}(\omega) = - \frac{1}{(2\pi)^2} \frac{{\sigma}_{(\mathrm{H})}(\omega)}{{\sigma}_{(\mathrm{L})}^{2}(\omega) + {\sigma}_{(\mathrm{H})}^2(\omega)} \; ,  
  \end{eqnarray}
which we calculated by using the maps in \eqref{Eq:Mapping}. If $W=0$ in the original theory then to compute the $ST^{K}$ transformed conductivites we set ${\sigma}_{(\mathrm{H})} = K/(2\pi)$. Similarly to compute the $T^{L}ST^{K}$ conductivities from those of the $ST^{K}$ we need only shift the Hall conductivity by $L/(2\pi)$. As a limiting case of the $ST^{K}$ transformed system notice that when $K \rightarrow 0$ we have $(2\pi)^2 {\sigma}_{(\mathrm{L})}(\omega) = 1/\sigma^{*}_{(\mathrm{L})}(\omega)$ and ${\sigma}_{(\mathrm{H})}(\omega) = \sigma^{*}_{(\mathrm{H})}(\omega) = 0$. We have analytically calculated the values of the DC conductivities for our system and they are given in \eqref{Eq:CTCLd0_norm} for the normally quantised system, \eqref{Eq:LowMomentumDataAlt} for the $S$ transformed system and \eqref{Eq:STnconductivities} for the $ST^{K}$ transformed system.}

{\ Now we consider small frequencies and small but non-zero momentum. Current conservation and Einstein's relation indicate that the longitudinal part of the Green's functions for current-current correlators have the form
  \begin{eqnarray}
    {C}_{(\mathrm{L})} =  \frac{\tilde{k} {\sigma_{(\mathrm{L})}}}{i \tilde{\omega} - {\tilde{D}} \tilde{k}^2}  \; , \qquad C^{*}_{(\mathrm{L})} =\frac{\tilde{k} \sigma^{*}_{(\mathrm{L})}}{i \tilde{\omega} - \tilde{D}^{*} \tilde{k}^2}  
		  = \frac{{C}_{(\mathrm{L})}}{(2\pi)^2 \left( {C}_{(\mathrm{T})} {C}_{(\mathrm{L})} + {W}^2 \right)} \; . 
  \end{eqnarray}
These relationships can be implied simply by considering a plane wave solution to the charge equation where $\omega \sim k^2, k \ll 1$. The final equality for $\tilde{C}^{*}$ simply repeats the relationship of \eqref{Eq:Mapping}. Using the definition of ${D}$ and $D^{*}$ we can eliminate ${C}_{(\mathrm{T})}$ and $C^{*}_{(\mathrm{T})}$ from our two-point functions. In particular, eliminating ${C}_{(\mathrm{T})}$, we find
  \begin{eqnarray}
   \label{Eq:AlternateDiffGeneralFormula}
     {C}_{(\mathrm{T})}
   = \frac{1}{\tilde{k} {\sigma}_{(\mathrm{L})}} \left[ i \tilde{\omega} - {H} \tilde{k}^2 \right] \;,\; 
     {H} 
   = \tilde{D}^{*} + \frac{{W}^2}{{\sigma}_{(\mathrm{L})}^2} \left( \tilde{D}^{*} - {\tilde{D}} \right) \; \mathrm{or} \;
     \tilde{D}^{*}
   = \frac{{H} + {\tilde{D}}\left(\frac{{W}}{{\sigma}_{(\mathrm{L})}}\right)^2}{1 + \left(\frac{{W}}{{\sigma}_{(\mathrm{L})}} \right)^2} \; , \; \; 
  \end{eqnarray}
where ${H}$ is some number which we can obtain from the normal quantisation two-point function and it is not to be confused with a diffusion constant. It is clear that, because we are working in a small frequency and momentum regime, only the constant part of ${W}$, equal to the DC Hall conductivity ${\sigma}_{(\mathrm{H})}$, can enter into defining $\tilde{D}^{*}$.}

{\ So far we have simply eliminated $C_{(T)}^{(*)}$ from the correlators at low frequency and momentum. This exercise is particularly useful for understanding the effect of an $ST^{K}$ transformation on the normally quantised system. We remind ourselves that adding a Chern-Simon's term to the initial theory only affects ${W}$ \eqref{Eq:TnTransformation}. As ${C}_{(\mathrm{T})}$ (and subsequently ${H}$) in \eqref{Eq:AlternateDiffGeneralFormula} is independent of performing the $T^{K}$-transformation but $W$ is shifted the dual diffusion must also shift to compensate for the change. In particular the new diffusion constant, denoted $D_{n}$, is given by
  \begin{eqnarray}
    \tilde{D}^{*} 
      \rightarrow \tilde{D}_{n} = \frac{\tilde{D}^{*} \left( 1 +\left(\frac{{\sigma}_{(\mathrm{H})}}{{\sigma}_{(\mathrm{L})}}\right)^2 \right) - {\tilde{D}} \left(\frac{{\sigma}_{(\mathrm{H})}}{{\sigma}_{(\mathrm{L})}}\right)^2 \left( 1 -\left(\frac{2 \pi {\sigma}_{(\mathrm{H})}+ K}{2 \pi {\sigma}_{(\mathrm{H})}}\right)^2 \right)}{1 + \left(\frac{2 \pi {\sigma}_{(\mathrm{H})}+K}{2 \pi {\sigma}_{(\mathrm{L})}} \right)^2} \; ,
  \end{eqnarray}
where $n = \frac{K}{2 \pi \mathcal{N}_{5}}$. There are several interesting limits of this expression. As a consistency check notice that for $K \rightarrow \infty$ we recover ${\tilde{D}}$ while for $K=0$ we find $\tilde{D}^{*}$. If $\tilde{D}^{*} = {\tilde{D}}$, for example when the theory is $S$-duality invariant, then $W$ and $K$ drop out of the expression and it reduces to an $K$-independent constant ${\tilde{D}}$. Finally, if ${\sigma}_{(\mathrm{H})}=0$ for our theory prior to acting with $ST^{K}$ we find
  \begin{eqnarray}
   \tilde{D}_{n} = \frac{\tilde{D}^{*} + {\tilde{D}} \left(\frac{K}{2 \pi {\sigma}_{(\mathrm{L})}} \right)^2}{1 + \left(\frac{K}{2 \pi {\sigma}_{(\mathrm{L})}} \right)^2} \; .
  \end{eqnarray}
}

{\ As an application of the maps in \eqref{Eq:AlternateDiffGeneralFormula} we shall now outline the calculation of the alternate quantisation diffusion constant from the transverse part of the normal quantisation two-point function. Letting $\omega \sim k^2, k \rightarrow \epsilon k, \epsilon \ll 1 $ we find that the near horizon expansion of the $a_{y}$ field is
  \begin{eqnarray}
   a_{y}(r) &=& a_{y}^{\mathrm{near}} \left[ 1 - i \epsilon^2 \frac{\left(\left(5 d^2 +  \left( \pi \ell T \right)^4 \right) \omega + 2 i \ell \left( \pi T \ell\right)^3 k^2 \right)}
								  {8 \left( \pi T \ell \right)^2 \left(d^2 + \left(\pi T \ell \right)^4\right)} \left(r - \pi T \ell^2 \right) \right. \nonumber \\
	     &\;& \left. \hphantom{ a_{y}^{\mathrm{near}} \left[ \right. } \vphantom{- i \epsilon^2 \frac{\omega}{4 \pi T} \log\left( \frac{r- \pi T \ell^2}{\pi T \ell^2} \right)} 
		  - i \frac{\epsilon^2 \omega}{4 \pi T} \ln \left(r-\pi T  \ell^2 \right)
		  + \mathcal{O}\left( (r - \pi T \ell^2)^2, (r - \pi T \ell^2) \ln (r - \pi T \ell^2), \epsilon^4 \right)  \right] \nonumber
  \end{eqnarray}
while at the boundary we identify
  \begin{eqnarray}
   a_{y}(r) = a_{y}^{(0)} + \left\langle J_{y} \right\rangle \frac{\ell}{r} + \mathcal{O}^{2}\left( \frac{\ell}{r} \right) \; .
  \end{eqnarray}
We expand the $a_{y}(r)$ equation of motion in a power series in small frequencies and momenta and solve the resultant equation of motion order by order matching to the near horizon expansion. Expanding the subsequent expression for $a_{y}(r)$ near the boundary allows us to identify the two-point function as
  \begin{eqnarray}
   \left\langle J_{y}(p) J_{y}(-p) \right\rangle &=& \mathcal{N}_{5} k \left[ 
   \frac{i \omega - \left(\frac{\pi T \ell}{\sqrt{d^2 + \left(\pi T \ell\right)^4}} \;_{2}F_{1}\left[\frac{1}{4},\frac{1}{2},\frac{5}{4};-\frac{d^2 \ell^4}{r^4} \right]\right) k^2 \ell}
   {\frac{(\pi T \ell)^2}{\sqrt{d^2 + \left(\pi T \ell\right)^4}}\left( 1 + \frac{a_{y}^{(1,0)}}{a_{y}^{(0,0)}} \omega \ell + \frac{a_{y}^{(0,2)}}{a_{y}^{(0,0)}} k^2 \ell^2 \right) k} \right] \; ,
  \end{eqnarray}
where $a_{y}^{(1,0)}/ a_{y}^{(0,0)}$ and $a_{y}^{(0,2)} / a_{y}^{(0,0)}$ are $\omega$ and $k$ independent constants that we will have no use for in the paper so we do not record them. They can be solved for from the near horizon boundary conditions. Given that $W=0$ we use \eqref{Eq:AlternateDiffGeneralFormula} to identify
  \begin{eqnarray}
    \label{Eq:LowMomentumDataAlt}
    \sigma^{*}_{(L)} = \frac{1}{(2\pi)^2 \mathcal{N}_{5} \sqrt{1+ d^2/(\pi T \ell)^4}} \;, \; \;
    \tilde{D}^{*} = \frac{(\pi T \ell)^2}{\sqrt{d^2 + \left(\pi T \ell\right)^4}} \;_{2}F_{1}\left[\frac{1}{4},\frac{1}{2},\frac{5}{4};-\frac{d^2}{\left(\pi T \ell\right)^4} \right]  \; , \; \;
  \end{eqnarray}
with the alternate Hall conductivity vanishing.}

\section{Normal quantisation data}
\label{normalquantisation}

\begin{figure}
 \centering
 \begin{subfigure}
  \centering
  \includegraphics[width=0.45\textwidth]{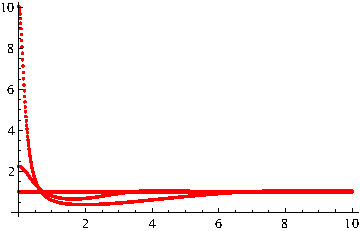}
 \end{subfigure} \hskip+3\unitlength
 \begin{subfigure}
  \centering
  \includegraphics[width=0.45\textwidth]{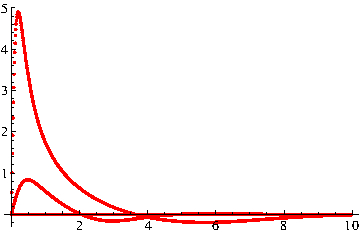}
 \end{subfigure}
 \begin{picture}(100,0)
  \put(16,36){\makebox(0,0){$\Re[{\sigma}_{(\mathrm{L})}(\tilde{\omega})],\; \Re[{\sigma}_{(\mathrm{T})}(\tilde{\omega})]$}}
  \put(65,35){\makebox(0,0){$\Im[{\sigma}_{(\mathrm{L})}(\tilde{\omega})],\; \Im[{\sigma}_{(\mathrm{T})}(\tilde{\omega})]$}}
  \put(49,7.5){\makebox(0,0){$\tilde{\omega}$}}
  \put(99,7.5){\makebox(0,0){$\tilde{\omega}$}}
 \end{picture}
 \vskip-2em
 \caption{Figures comparing analytic expressions for ${\tilde{\sigma}}_{(\mathrm{L})}$ and ${\tilde{\sigma}}_{(\mathrm{T})}$ associated with spatially homogeneous gauge perturbations in a thermal strongly coupled $(2+1)$-dimensional field theory with background charge $\tilde{d}$. Red dots are numerical data. \textbf{Left:} The frequency dependent real parts of ${\tilde{\sigma}}_{(\mathrm{L})}$ and ${\tilde{\sigma}}_{(\mathrm{T})}$. The horizontal line is $\tilde{d}=0$ while along the vertical axis the highest line is $\tilde{d}=10$ while the middle line 
is $\tilde{d}=2$. \textbf{Right:} The frequency dependent imaginary parts 
of ${\sigma}_{(\mathrm{L})}$ and ${\sigma}_{(\mathrm{T})}$. The line with the greatest range is $\tilde{d}=10$, that along the axis is $\tilde{d}=0$ while the remaining line is $\tilde{d}=2$.}
 \label{Fig:NonzerodDiffusionConstants_norm}
\end{figure}
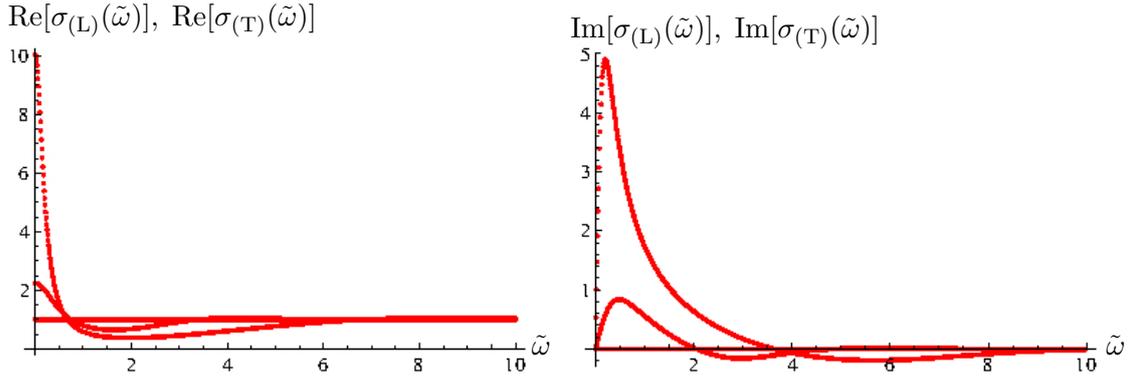

{\ The results in this section are well known and we simply discuss them for the purposes of comparison. Consider the normally quantised system with non-zero background charge density $d$ at non-zero temperature. We can probe the system by perturbing the $U(1)$ charge vector about it's background value of $\langle J^{t} \rangle \propto \tilde{d}$. Let the perturbation be spatially homogeneous then the DC conductivity calculated from the retarded two-point function of two such charge currents is given by
  \begin{eqnarray}
   \label{Eq:CTCLd0_norm}
   \sigma_{(\mathrm{L})} = \mathcal{N}_{5} \sqrt{1+\tilde{d}^2} 
  \end{eqnarray}
In fig~\ref{Fig:NonzerodDiffusionConstants_norm} we display the AC conductivites which should be compared with fig.~\ref{Fig:NonzerodConductivitiesvariousn}.}
  
{\ Now consider spatially inhomogeneous perturbations. At $\tilde{d}=0$ the lowest lying poles are given exactly by the $\tilde{d}=0$ subfigure of fig~\ref{fig:dualdiffusion}. Increasing $\tilde{d}$ to any non-zero value it is known \cite{Brattan:2012nb} that for small enough $\tilde{k}$ the picture remains similar to $\tilde{d}=0$ where two imaginary modes combine into a complex mode that becomes light-like at large $\tilde{k}$. One of these imaginary modes is diffusive for sufficiently small $\tilde{k}$ with a diffusion constant
  \begin{eqnarray}
   \label{Eq:NormDiff}
   {\tilde{D}} = \frac{1}{2 \tilde{d}^2} \left[ 1 - \sqrt{1 + \tilde{d}^2} \;_{2}F_{1}\left[-\frac{3}{4},\frac{1}{2},\frac{1}{4};-\tilde{d}^2\right] \right] \;.
  \end{eqnarray}
The cross-over from diffusive to collisionless behaviour occurs for progressively lower values of $\tilde{k}$ as $\tilde{d}$ increases. The absolute value of frequency where this cross-over happens is given by $\left| \omega \right| \sim 0.3 ( \pi T / \sqrt{d} )^2$ for sufficiently large $\tilde{d}$ \cite{Brattan:2012nb}. Further for sufficiently large $\tilde{d}$ a new regime appears as a remenant of a sound-like mode which dominates the late time behaviour of the system at zero temperature and non-zero density \cite{Brattan:2012nb}. This mode has the well known \cite{Karch:2008fa} dispersion relation:
  \begin{eqnarray}
   \omega(k) = \pm \frac{1}{\sqrt{2}} k - i \frac{d^{1/2}}{4 \mu_{T=0}} k^2 + \mathcal{O}^3\left(k\right) \; , \qquad  \mu_{T=0} = \frac{\Gamma\left(1/4\right)\Gamma\left(5/4\right)}{\Gamma(1/2)} \, d^{1/2} \; . 
  \end{eqnarray}
The presence of this new regime can be identified by an additional kink in the real part of the dispersion relation $\tilde{\omega}(\tilde{k})$ with $\tilde{k}$ is real.}
  
\bibliographystyle{JHEP}
\bibliography{alt_quant}

\end{document}